\documentclass[fleqn,usenatbib]{mnras}

\pdfoutput=1   

\usepackage[T1]{fontenc}
\usepackage{ae,aecompl}

\usepackage{graphicx}	
\usepackage{amsmath}	
\usepackage{amssymb}	
\usepackage{multirow}
\usepackage[usenames, dvipsnames]{color}
\usepackage{lscape}
\usepackage{hyperref}

\usepackage[normalem]{ulem} 

\title[Mid-IR properties of AT20G compact AGNs]{WISE Mid-Infrared Properties of compact Active Galactic Nuclei selected from the high-radio-frequency AT20G Survey}

\author[R. Chhetri et al.]{
R. Chhetri,$^{1,2}$\thanks{E-mail: rzn.chhetri@gmail.com}
A. Kimball,$^{3}$
R. D. Ekers,$^{1,4}$
E. K. Mahony$^{4}$,
E. M. Sadler$^{4, 5}$,
\newauthor ~and T. Jarrett$^{5}$
\\
$^{1}$ International Centre for Radio Astronomy Research, Curtin University, GPO Box U1987, Perth, WA 6845, Australia\\
$^{2}$ CSIRO Astronomy and Space Science, 26 Dick Perry Avenue, Kensington, WA 6151, Australia\\
$^{3}$ National Radio Astronomy Observatory, 1003 Lopezville Rd, Socorro, NM, 87801, USA \\
$^{4}$ CSIRO Astronomy and Space Science, PO Box 76, Epping, NSW 1710, Australia\\
$^{5}$ Sydney Institute for Astronomy, School of Physics, University of Sydney, NSW 2006, Australia \\
$^{6}$ Department of Astronomy, University of Cape Town, Private Bag X3, Rondebosch, 7701, South Africa\\
}
\date{Accepted XXX. Received YYY; in original form ZZZ}

\pubyear{2018}

\begin{document}

\label{firstpage}
\pagerange{\pageref{firstpage}--\pageref{lastpage}}
\maketitle

\begin{abstract}

Past studies of compact active galactic nuclei (AGNs), the dominant population at high radio frequencies, selected them using flat spectral index criteria. This biases the sample due to the steepening of AGN spectra at high radio frequencies. We improve upon this by selecting 3610 compact AGNs using their angular size information ($\sim$0.15 arcsec scale) from the Australia Telescope 20 GHz (AT20G) high-angular-resolution catalogue. We cross-match these against the Wide-field Infrared Survey Explorer AllWISE catalogue and present a catalogue with 3300 (91\%) matches, 91 (3\%) rejects and 219 (6\%) nondetections that are excellent high redshift candidates. Of the matched compact AGNs, 92\% exhibit QSO mid-infrared colours (W1-W2>0.5). Therefore, our sample of high frequency compact sources has a very high rate of identification with mid-infrared QSOs. We find counterparts for 88\% of 387 compact steep-spectrum (CSS) sources in the AT20G survey, 82\%$\pm$5\% of which exhibit QSO mid-infrared colours and have moderate redshifts (z$_{median}$=0.82), while those dominated by host galaxy colours in mid-infrared have lower redshifts (z$_{median}$=0.13). The latter classified into late- and early-type galaxies using their mid-infrared colours shows a majority (68\%$\pm$4\%) have colours characteristic of late-type galaxies. Thus, we find that a larger fraction of these CSS sources are embedded in hosts with higher gas densities than average early-type galaxies. We compare mid-infrared colours of our AGNs against those reported for AGNs primarily selected using non-radio techniques. This shows that mid-infrared SED of high frequency selected compact radio AGN is comparatively less red, possibly due to contributions from their hosts.

\end{abstract}



\section{Introduction}

At frequencies above a gigahertz, the extragalactic radio source population is increasingly dominated by objects from a flat spectrum population as expected for a high frequency sample \citep{ Dezotti2005, Massardi2011, Chhetri2013}. The flat-spectrum FR 0 \citep{Baldi_2015A&A...576A..38B} population comprises the compact (at sub-arcsecond angular scales) radio sources with radio emission from the vicinity ($\sim$ kiloparsec scales) of supermassive black holes at the centers of galaxies. The population includes a mixture of relativistically beamed multiple synchrotron absorbed components \citep{Kellermann1969_apjl_155} and the young compact GHz peaked spectrum sources \citep{O'Dea1998}.  

Since there are insufficient angular size measurements to identify such compact sources in large numbers from current surveys, the classifications of sources in this population have traditionally been made on the basis of source spectral index \citep[e.g.][]{Wall1977__74}. A flat spectral index has been used to select compact active galactic nuclei (AGNs) and a steep spectral index to select the extended radio galaxies (both~Fanaroff-Riley class I and II \citep[FR\,I, FR\,II respectively,][]{Fanaroff1974} extended sources).  There is, however, an important complication in this approach to select clean samples of AGNs. The spectra of compact AGNs also steepens at rest frame frequencies $\lesssim$30\,GHz \citep{Massardi2011_mnras_415, Chhetri2012}. Thus, the radio spectra of compact AGNs can appear steep at lower observed frequencies due to the effects of redshift and this will introduce a redshift bias in a spectral index selected sample. For example, AGNs at redshift above 0.5 start to show steepening of their spectra below 20\,GHz, and 6.7\%$\pm$0.5\% of flat-spectrum sources at 20\,GHz fall out of the flat-spectrum criteria by 5\,GHz \citep[related discussions are in][]{Chhetri2013a}. An approach based on angular sizes to select compact AGNs, therefore, is more desirable. This is made possible for a very wide area survey by the Australia Telescope 20 GHz (AT20G) high-angular-resolution catalogue \citep{Chhetri2013}, which covers the entire Southen hemisphere. The catalogue, based on the AT20G survey \citep{Murphy2010}, was made at high radio frequencies of 20\,GHz and identifies compact objects at $\sim$0.15 arcsecond angular size scales, enabling studies of source population of compact AGNs from a large-area survey.

For the compact AGN sample we also have an important practical advantage since unambiguous identification at different wavelengths is possible. This is much more difficult for the extended radio sources which are often much larger than their host galaxies. With this objective in mind we have made a cross-match of radio-selected compact AGNs with the Wide-field Infrared Survey Explorer (WISE) catalogue \citep{Wright2010}. The WISE catalogue of celestial sources with mid-infrared (MIR) emission has been used widely to classify source populations and to gain insights into their nature. For example, the colours in the WISE observing bands of 4.6 and 12 microns have been used to make a distinction between populations of elliptical and spiral galaxies \citep[e.g.,][]{Wright2010, Sadler2014, Jarrett_2017ApJ...836..182J}, while colours from 3.4 and 4.6 (and 12) microns have been used to separate AGNs from galaxies \citep[e.g.][]{Jarrett2011,Stern2012, Mateos2012, Satyapal2014ApJ...784..113S, Hickox2017}. In this work, we present the identifications of MIR counterparts of compact radio AGNs selected from a high radio frequency catalogue, and an investigation of their MIR properties.

The layout of this paper is as follows. In Section 2, we describe the parent surveys and identify the parent AGN population. In Section 3, we provide details of our cross-match between the AT20G and WISE surveys. In Section 4, we present the resulting AT20G-WISE catalogue. In Section 5, we analyze the results and discuss their implications. We summarize our main findings in Section 6. Further details related to the cross-match work, details of relevant flags used in the WISE catalogue, and some specific sources are provided in the appendices.

Throughout this paper we have assumed a lambda cold dark matter cosmology with $\Omega_m$ = 0.27, $\Omega_{\lambda}$ = 0.73 and $H_0$ = 71 km\,$s^{-1}$\,Mpc \citep{Larson2011}. For radio spectral-index measurements, we use the definition $S_{\nu}\,\propto\,\nu^{\alpha}$, where S, $\nu$ and $\alpha$ are flux density, frequency and spectral index, respectively. We use the symbol $\alpha^{\nu_1}_{\nu_2}$ to denote spectral index calculated between frequencies $\nu_1$ and $\nu_2$ in GHz.

\section{The parent surveys}
\label{6kvis}

\subsection{The AT20G high-angular-resolution catalogue}
This work uses compact sources at sub-arcsecond angular scales identified in the AT20G high-angular-resolution catalogue \citep{Chhetri2013}. 
The compact sources are identified in the AT20G survey using the complex visibility amplitude ratio between the long ($\sim$4500m) and short baselines of the Australia Telescope Compact Array (ATCA) at 20 GHz. Thus, sources with this ratio (termed the 6-km visibility\footnote{Because the long baselines visibilities are obtained using the only fixed ATCA antenna, called the ``6-km antenna''.}) $\geq$ 0.86 are compact at 0.15 arcsec angular scales. 351 (6 per cent) out of the 5890 objects in the AT20G survey do not have this measurement due to unavailability of the 6-km antenna during some observations (e.g. due to maintenance). For this reason, these sources are omitted from this work. Since the unavailability of the 6-km antenna is not related to astrophysical properties of the sources in the sky, the omission of sources lacking measured 6-km visibilities does not introduce any bias. 

The high-angular-resolution catalogue has the same sky coverage (all Southern hemisphere) and well known completeness and reliability as that of the AT20G catalogue. The AT20G survey is a flux-density-limited survey, with a limit of 40 mJy/beam at 20 GHz, and has a Galactic cut-off of |b|=1.5 degrees.

\begin{center}
\begin{figure}
\includegraphics[scale=0.83, angle=0]{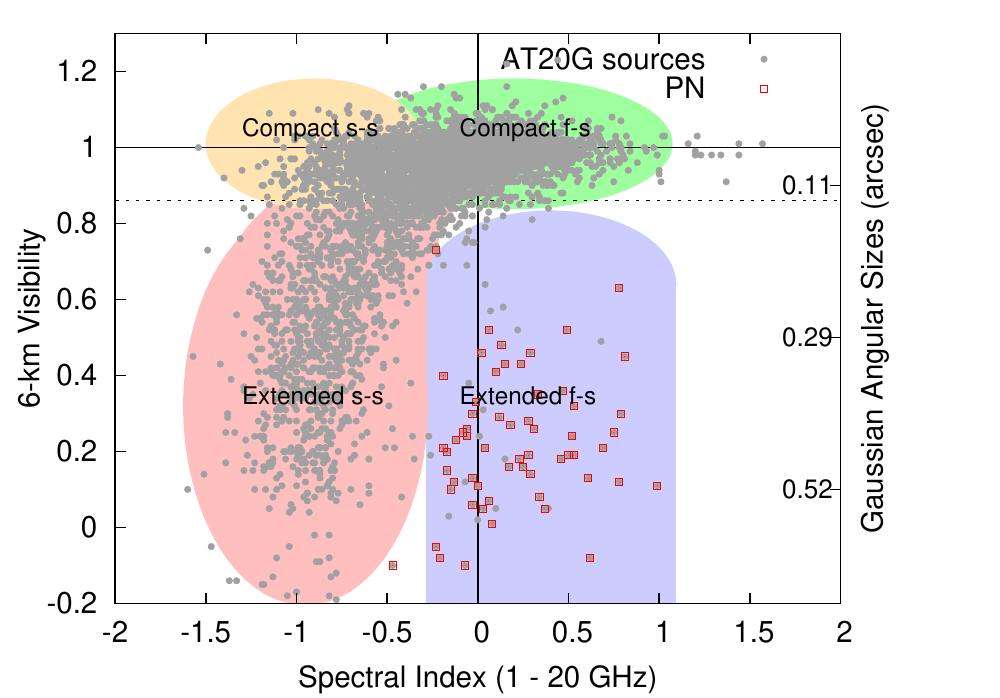}
\caption{Visibility-spectra diagnostic plot \citep[modified from][]{Chhetri2015} identifying regions occupied by different galactic/extra-galactic populations in the AT20G survey. Green, yellow, pink and blue highlighted areas are, respectively, occupied by compact flat-spectrum sources, compact steep-spectrum sources, extended steep-spectrum sources and an intresting population that are ``extended'' and exhibit flat- or rising-spectra. Sources that have been identified with galactic or extragalactic thermal sources are highlighted with red squares, and are removed from this analysis. Spectral index is shown in X-axis, and 6-km visibility on the Y-axis, with corresponding angular scales using a Gaussian model shown in Y-axis on the right hand side.}
\label{Fig:visib-spectra}
\end{figure}
\end{center}

The brightness sensitivity limit of the AT20G survey means that the radio emission detected for sources (including for the small number of low redshift galaxies detected in the survey) is due to an active nucleus rather than star formation processes \citep{Murphy2010}. Therefore, essentially all extragalactic AT20G sources are AGNs.
The focus of this paper is the population of parsec-scale compact active cores, not the highly extended radio galaxies. We also note that the very high surface density of sources in the WISE catalogue makes it difficult to unambiguously cross-match with the extended radio sources, and we do not attempt to do so in this work. The majority of AT20G sources are compact as shown in \autoref{Fig:visib-spectra}. Our visibility criterion easily separates the compact active nuclei of galaxies from the extended radio galaxies that exhibit steep radio spectra. We use the term active galactic nuclei (AGNs) to denote these compact objects. There are 4259 compact objects, which is 77 per cent of the 5539 sources with 6-km visibilities in the AT20G high-angular-resolution catalogue.

\subsubsection{Effect of redshift on angular sizes}

The changing effect of redshift on angular size of a source means that our use of 6-km visibility limit of 0.86 probes different linear sizes for different redshift values. This effect is discussed in \cite{Chhetri2012} and \cite{Chhetri2013} (see their figure 2 and 3, respectively). The limit of 0.86 corresponds to a linear size of $\sim$1\,kpc for a large range of redshifts (0.7<z<5). 
For z$<$0.7 the linear scales probed decrease with z to $\sim$0.2\,kpc at z = 0.1.
Since there is a large difference between the linear sizes of extended radio galaxies (with sizes ranging from few hundred kpc to megaparsec scales; \cite{Miley1980ARA&A..18..165M}) and the compact flat-spectrum cores (compact at kpc scales \citep{Baldi_2015A&A...576A..38B}, and as small as pc scale \citep{Miley1980ARA&A..18..165M}), we expect that the linear sizes probed by our cut-off do not significantly bias against selection of flat-spectrum cores. However, for sources with slightly extended structures (e.g. gigahertz peaked spectrum sources and compact steep spectrum sources), our sample will progressively miss kpc scale sources below z<0.7.

\subsection{The WISE survey}
The NASA WISE satellite mapped the sky at four mid-infrared wavelength bands centered at 3.4, 4.6, 12 and 22 microns between 7 January and 6 August 2010. These bands are referred to as W1, W2, W3 and W4 bands, and have angular resolutions (full width at half maximum of the major axes of the point spread function) of 6.1, 6.8, 7.4 and 12 arcseconds respectively. 
This work uses the AllWISE data release\footnote{http://wise2.ipac.caltech.edu/docs/release/allwise/expsup/} that contains photometric and positional information for over 700 million objects detected in the WISE images. 
The astrometric precision for quasars in the AllWISE survey catalogue is better than 0.15 arcsec. 
The catalogue is >95 per cent complete for sources with W1<17.1\,mag, W2<15.7\,mag, W3<11.5\,mag and W4<7.7\,mag. All reported magnitudes are in the Vega system.

\section{The AT20G - WISE Cross-Match Details}\label{Sec:xmatch}

\subsection{The AT20G parent sample of compact sources (<0.15 arcsec)}

The WISE survey is very sensitive to galactic thermal sources. Hence, in order to avoid possible incorrect matches of our compact source sample with galactic sources we excluded sources with Galactic latitude $|b|<10^{\circ}$ from the AT20G high-angular-resolution catalogue. The remaining 3610 compact AGNs form our parent sample for cross-matching. Of these, 91 nearby AT20G sources have been included in \cite{Sadler2014} using the WISE All-Sky catalogue \citep{Cutri2012}. The All-Sky catalogue has been updated with the AllWISE catalogue. Therefore, for these sources we used the WISE counterparts provided in \cite{Sadler2014} but obtained updated AllWISE photometry from the WISE consortium (T.~Jarrett). We then cross-matched the remaining 3519 AT20G compact AGNs against the AllWISE catalogue. Of these, 2844 (79\%$\pm$2\%) sources have optical identifications as published in \cite{Mahony2011}. These identifications are based on the digitized UK Schmidt Telescope survey in the SuperCOSMOS database \citep{Hambly2001MNRAS.326.1279H}, and redshift information were found from either the 6dF Galaxy survey or the literature.

\subsection{Cross-match methodology}
We used the radio position---or optical position when available---for the 3519 sources to make cross-matches with the WISE catalogue. The optical position, when available, was given higher priority since its use improved cross-match for a small fraction (approximately 3 per cent) of AT20G compact sources with large errors in the radio position (see \autoref{Fig:posOffsets}). The cross-match was made using a radial search around our source position. 

In order to identify the best cross-match radius we looked at the distribution of the separation of WISE neighbours as a function of distance from the AT20G sources. This is shown in \autoref{Fig:astrometry} for all AT20G sources away from the Galactic plane, and it compares the equivalent distribution for random source positions. The random source positions were obtained by shifting (increasing) the positions of AT20G sources by 1 degree in Galactic longitude. The lower panel of \autoref{Fig:astrometry} suggests that 2.5 arcsec is a reasonable matching radius to choose so as to balance between completeness and reliability of the cross-matched sample. Using the formalism presented in \citet{Condon1975} and \citet{Condon1998}, we estimate the probability of false WISE match to AT20G sources within 2.5 arcsec radius to be 0.36 per cent. We used this search radius for our automatic matching criteria (see Section \ref{Sec:SingleCounterpart}). In order to find genuine counterparts that are farther than the optimum 2.5 arcsec radius, we extended our counterpart search to 7 arcsecond of our search positions, $\sim$3 times the optimum search radius, and identified or rejected matches (see Section \ref{subsec:crossmatch}). Here, the probability of false WISE match rises to 3 per cent. We made visual inspections of all sources whose counterparts are outside the 2.5 arcsec radius (see Sections \ref{Sec:SingleCounterpart_7asec} and \ref{Sec:MultipleCounterparts}). Visual inspections were made on WISE images overlaid with our radio source position and contours, and with optical contours when available. Beyond the radius of 7 arcsec from the search position, the probability of finding counterparts due to chance alignment becomes prohibitive.

\subsubsection{Deficiency in counterpart numbers}\label{Sec:Dearth}
\begin{center}
\begin{figure}
\includegraphics[scale=0.38, angle=0]{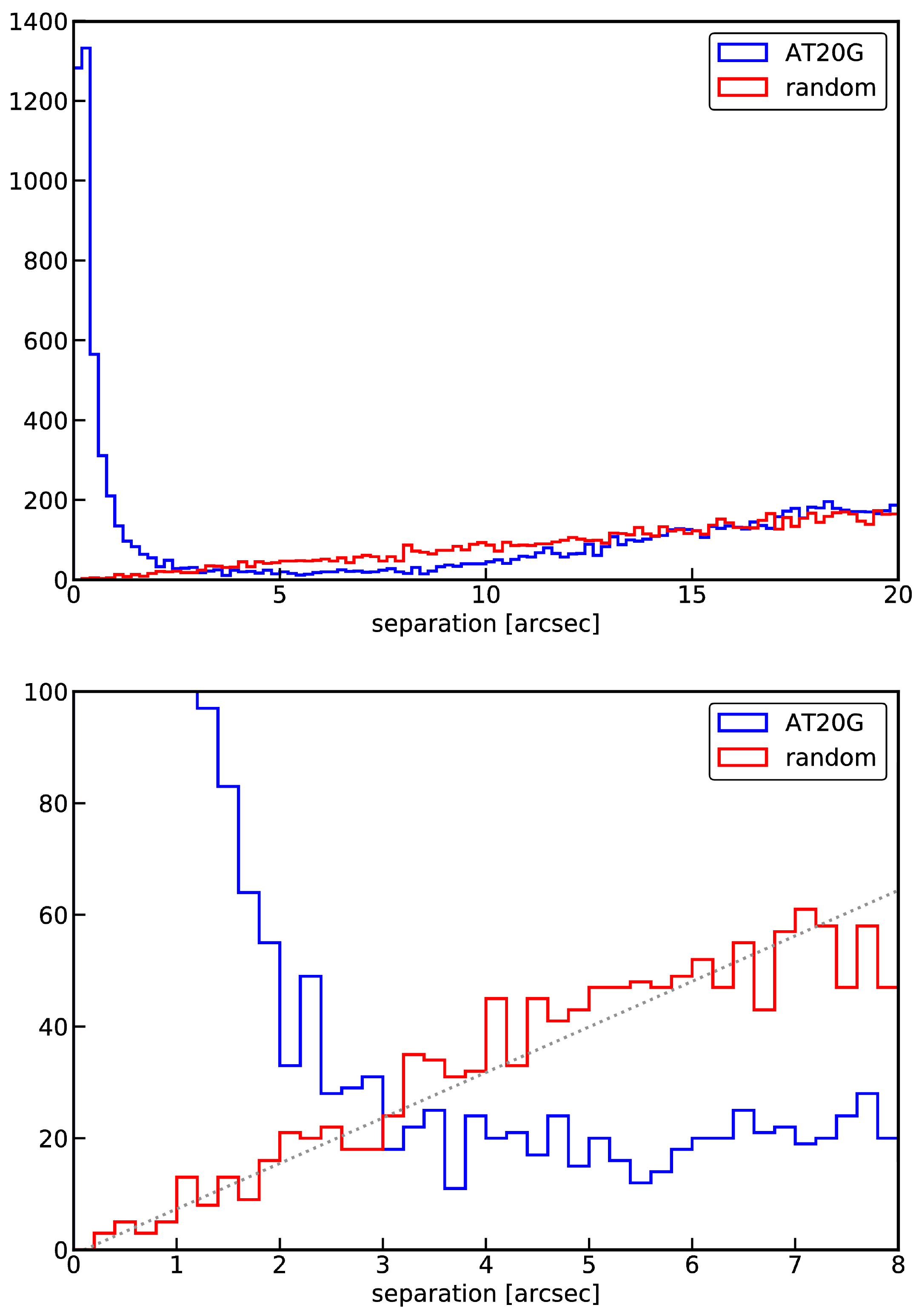}
\caption{Plots showing the comparative distributions of all WISE objects near AT20G sources (blue line) and those near random source positions (red line) as a function of search radius. The bottom frame shows that the probability of finding a cross-match by chance for a random position is very small within a 2.5 arcsec radius, compared to real AT20G source positions. The dotted line shows the linear fit to the distribution of chance cross-matches. A discussion of the dip in the number of WISE objects in the real source positions compared to that by chance is given in Section \ref{Sec:Dearth} and Appendix \ref{appendix:Dearth}.}
\label{Fig:astrometry}
\end{figure}
\end{center}

The result from a positional cross-match between two catalogues is the union of true cross-matches and any cross-matches due to chance alignments. In \autoref{Fig:astrometry} we see a curious deficiency in the number of all counterpart sources between $\sim$4 and $\sim$14 arcsec for cross-matches between WISE and AT20G positions when compared to cross-matches of WISE and random (test) positions. A similar deficiency was observed by \cite{Krawczyk2013} as seen in their figure 4.  
We investigated this oddity and found that pixels near strong sources in WISE images are affected by saturation. This effect renders weaker sources close to the strong sources in these images indistinguishable from the strong sources in the de-blending step. The distance to which this effect is expected to affect weak sources close to strong sources is, in practice, larger than $\sqrt{2}$ times the FWHM of the WISE beam\footnote{Private communications Ken Marsh, WISE Science Data Center Cognizant Engineer/Scientist.}. We made similar plots to test our understanding for weaker sources and resolved sources.  We found that the distribution of cross-matches for weaker sources as well as for resolved/extended sources in SDSS/WISE exhibit a very pronounced effect which supports our explanation for the deficiency in the number of sources seen in \autoref{Fig:astrometry}. The results for weaker and resolved/extended sources and a more detailed explanation are presented in Appendix \ref{appendix:Dearth}.

\subsection{Result of cross-match}
\begin{figure*}
\centering
\includegraphics[width=\textwidth, angle=0]{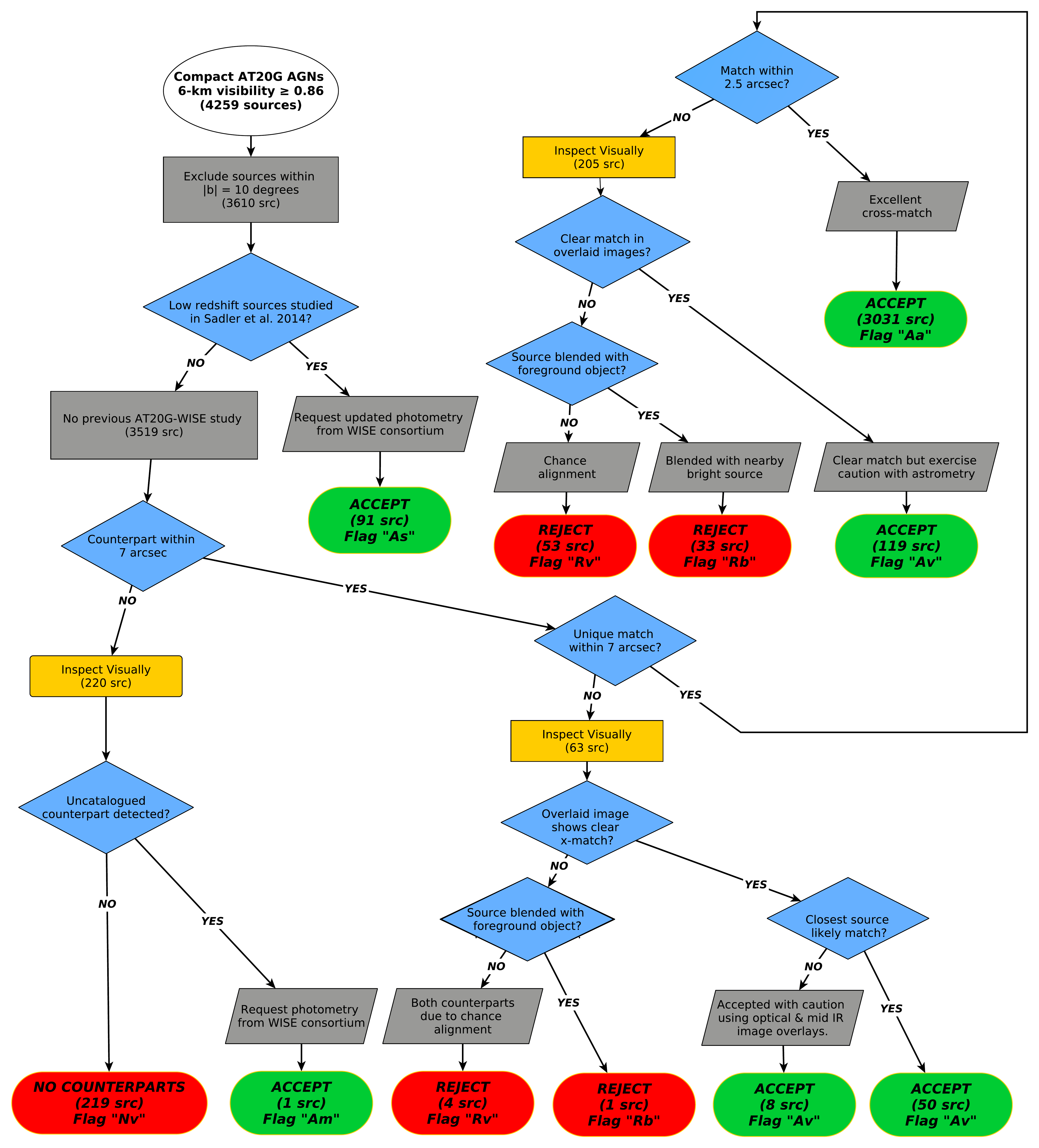}
\caption{Flowchart showing the steps used to confirm or reject WISE counterparts for AT20G compact AGNs.}
\label{Fig:flowchart}
\end{figure*}

{\label{subsec:crossmatch}}
All sources in the WISE survey found within the 7 arcsecond of our search positions were considered as potential counterparts to AT20G sources. These counterparts were then either selected automatically, selected as a counterpart after visual examination or rejected due to various reasons as shown in the flow chart in \autoref{Fig:flowchart}.

\subsubsection{Single counterpart - automatic acceptance}
\label{Sec:SingleCounterpart}
\begin{figure*}
\centering
\includegraphics[width=\textwidth, angle=0]{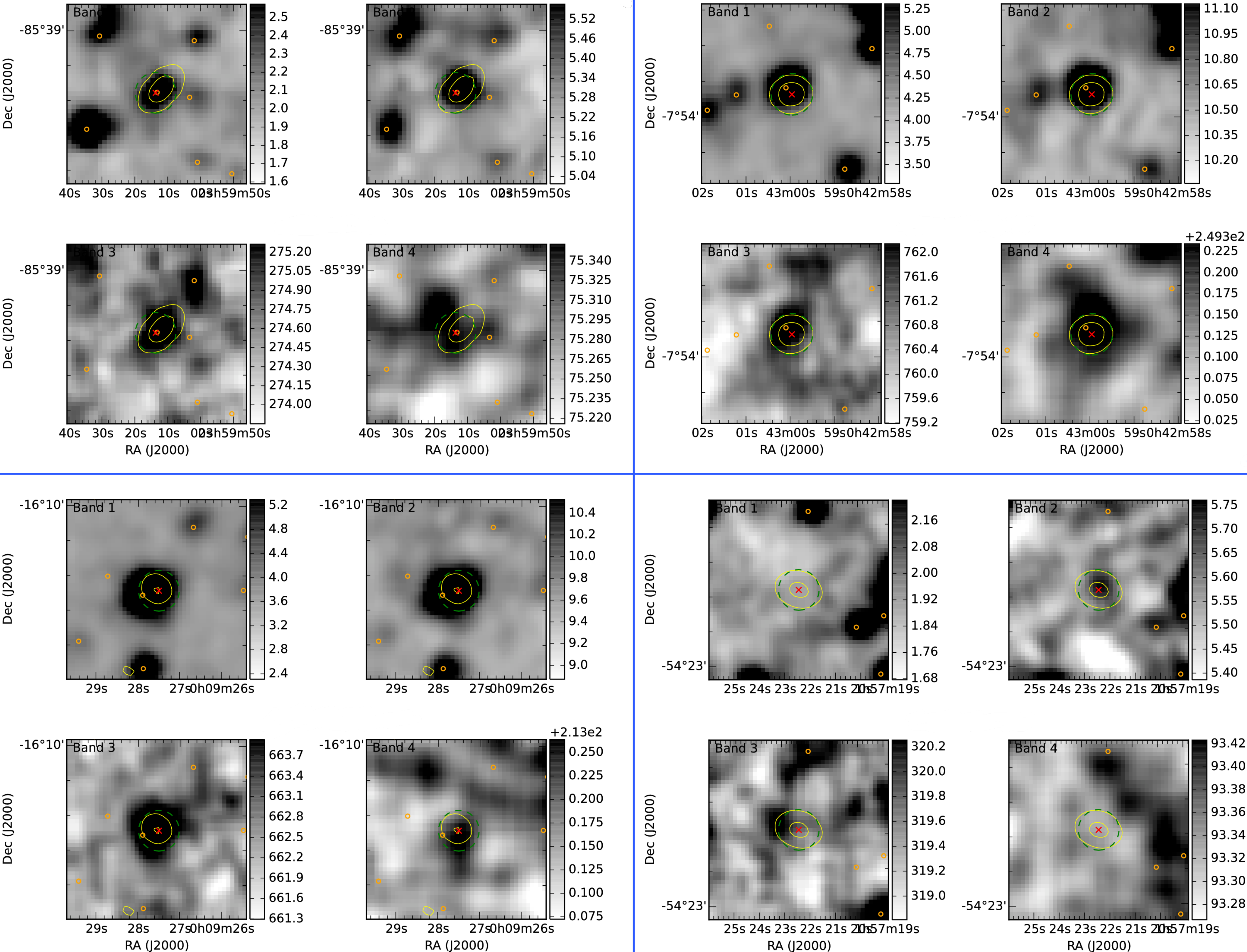}
\caption{Figures show examples for four typical scenarios encountered during our cross-matching steps. Each panel, separated by blue lines, shows images in all four WISE bands. The top left panel shows an example of a source whose cross-match was automatically (unambiguously) accepted. The top right panel shows an example of a source that was accepted after visual inspection of the overlaid images. The bottom left panel shows an example of a source blended with another (possibly foreground) object, hence the cross-match was rejected. The bottom right panel shows an example of a source that is a non-detection in WISE. In all figures, the red cross is the search position, yellow contours are from the AT20G survey, the dotted green circle is the 7 arcsec search radius and the orange circles are identifications listed in the AllWISE catalogue. }
\label{Fig:match_examples}
\end{figure*}

There are 3236 sources with unique counterparts within 7 arcsec radius. Of these, 3031 sources have unique counterparts within 2.5 arcsec. These were automatically selected as correct counterparts for the AT20G sources. A ``\textit{Aa}'' (accepted automatically) flag has been given to these sources in the catalogue (see example: the top left panel of \autoref{Fig:match_examples}).

\subsubsection{Single counterpart - visual inspection}
\label{Sec:SingleCounterpart_7asec}
We visually examined the WISE images for the remaining 205 sources with unique counterparts within 7 arcsec but outside 2.5 arcsec of our source positions (see the top right panel of \autoref{Fig:match_examples}). These inspections were made independently by three co-authors (A.\,K., E.\,K.\,M. and R.\,C.) and the final decision was made by collating the outcomes of these inspections. Since we expect our high frequency selected compact sources to be AGNs, we made identifications with the prior knowledge that AGNs are brighter in W3 and W4 compared to W1 and W2 in WISE (whereas Galactic objects tend to be brighter in W1 and W2 than in W3 and W4). As an example, if multiple potential matches were found within the 7 arcsec search radius the WISE source that appeared brighter in W3 and W4 was chosen as the correct counterpart even if the source was farther than other potential counterparts.

We accepted the cross-matches for 119 of these sources and a ``\textit{Av}'' (accepted after visual inspection) flag has been given to these sources in the catalogue. Some of the counterparts for these sources appear to be extended in either optical (when available) or WISE images. Some of these sources appear close to a strong source and suffer from low level blending in WISE images. Thus, we note that the astrometry of sources with \textit{Av} flag may be less reliable than those with \textit{Aa} flag. We note that 80 of the 91 sources studied in \cite{Sadler2014} would have the \textit{Av} flag since these are sources from the local universe that project larger angles in the sky.

Of the remaining 86 sources with unique possible counterparts within 7 arcsec, 53 sources appeared bright in W1 and W2 compared to other WISE bands. We concluded that they are associated with possible foreground objects due to alignment by chance and rejected these cross-matches. They are given the ``Rv'' (rejected after visual inspection) flag in our catalogue. 

The remaining 33 objects that were visually inspected, while brighter in W3 and W4, were blended with stronger foreground stars that are stronger in W1 and W2 bands. Thus, it was not possible to obtain their true WISE magnitudes. These have been given the \textit{Rb} flag in the catalogue. Since the rejected sources appear blended with unrelated foreground stars, their rejection from the catalogue should not introduce any bias. 

\begin{table*}
\centering
{
\begin{tabular}{lllllll}
\hline
\multicolumn{2}{c}{\textbf{Counterpart within}}	&	\multicolumn{2}{c}{\textbf{Unique within}}	& \multicolumn{2}{c}{\textbf{Accepted from}}	&\\
\textbf{2.5\arcsec}	& 	\textbf{7\arcsec}	&\textbf{2.5\arcsec}	& 	\textbf{7\arcsec}	& \textbf{This work} 	& \textbf{\cite{Sadler2014}}	& \textbf{Total}\\
\hline 
Yes	&	Yes	&	Yes	&	Yes	&	3031 	&	80	&	3111\\
No	&	Yes	&	N/A	&	Yes	&	119	&	4	&	123\\
Unknown 	&	Yes	&	Unknown	&	No	& 	58	& 	7	&	65\\
No	&	No	&	N/A &	N/A	&	1	&	0	&	1\\
\hline
\multicolumn{4}{c}{\textit{Total}}		&	3209	&	91	&	3300\\
\hline
\end{tabular}
}
\label{table:accepted}
\caption{We present the number of sources that were accepted via different paths in the flowchart shown in \autoref{Fig:flowchart}.}
\end{table*}

\subsubsection{Multiple counterparts - visual inspection}
\label{Sec:MultipleCounterparts}
There are 63 sources with multiple counterparts within 7 arcsec of our search position. \autoref{Fig:astrometry} indicates that this is approximately the number of objects for which we find one true and at least one other counterpart by chance. After visual inspection we accepted the closest counterpart for 50 sources. These objects are also given \textit{Av} flag in the catalogue. 
Eight more sources had complicated cross-matches but it appears that the counterpart with a slightly farther separation is the true counterpart. These 8 were also accepted with the ``Av" flag.

One of the remaining 5 sources with multiple counterparts appeared blended with a foreground object in WISE images. Therefore, this source was rejected and given a ``Rv'' flag. The remaining four sources had two potential counterparts, each within 7 arcsec. Visual inspections of overlaid images indicate that both potential counterparts are chance alignments and hence we rejected them (and gave them \textit{Rv} flag). These four sources are J052335-513830, J092902-182048, J190106-173840, J235025-022442.

\subsubsection{No counterpart}

Of the remaining 220 sources, we determined that 219 sources have no visible counterparts within 7 arcseconds in WISE images after visual inspection. These sources have been given ``Nv'' flag in the catalogue (see example: the bottom right panel of \autoref{Fig:match_examples}). For a small number of these sources, possible weak mid-infrared counterparts in the images appear to be heavily swamped by nearby very bright sources. One source shows a potential counterpart just outside the 7 arcsec radius, but it is blended with a stronger foreground source. The implications of these non-detections are discussed further in Section \ref{Sec:non_detections}.

AT20G source J022740-302604 shows a clear counterpart in the WISE images but it was not listed in the AllWISE catalogue. We requested WISE photometry for this source (private communications: T.~Jarrett) and have presented the accepted counterpart photometry in the catalogue (catalogue flag ``Am''). \\

Thus, from a parent sample of 3610 extragalactic compact sources in AT20G high-angular-resolution catalogue, we present WISE mid-infrared counterparts for 3300 (91.4 per cent) sources.  \autoref{Fig:posOffsets} shows the positional offset between our search position and WISE positions. We either did not find a counterpart in WISE or rejected the cross match for 310 (8.6 per cent) sources.

\subsubsection{Sources with large offsets in positions}
\begin{center}
\begin{figure}
\includegraphics[scale=0.48, angle=0]{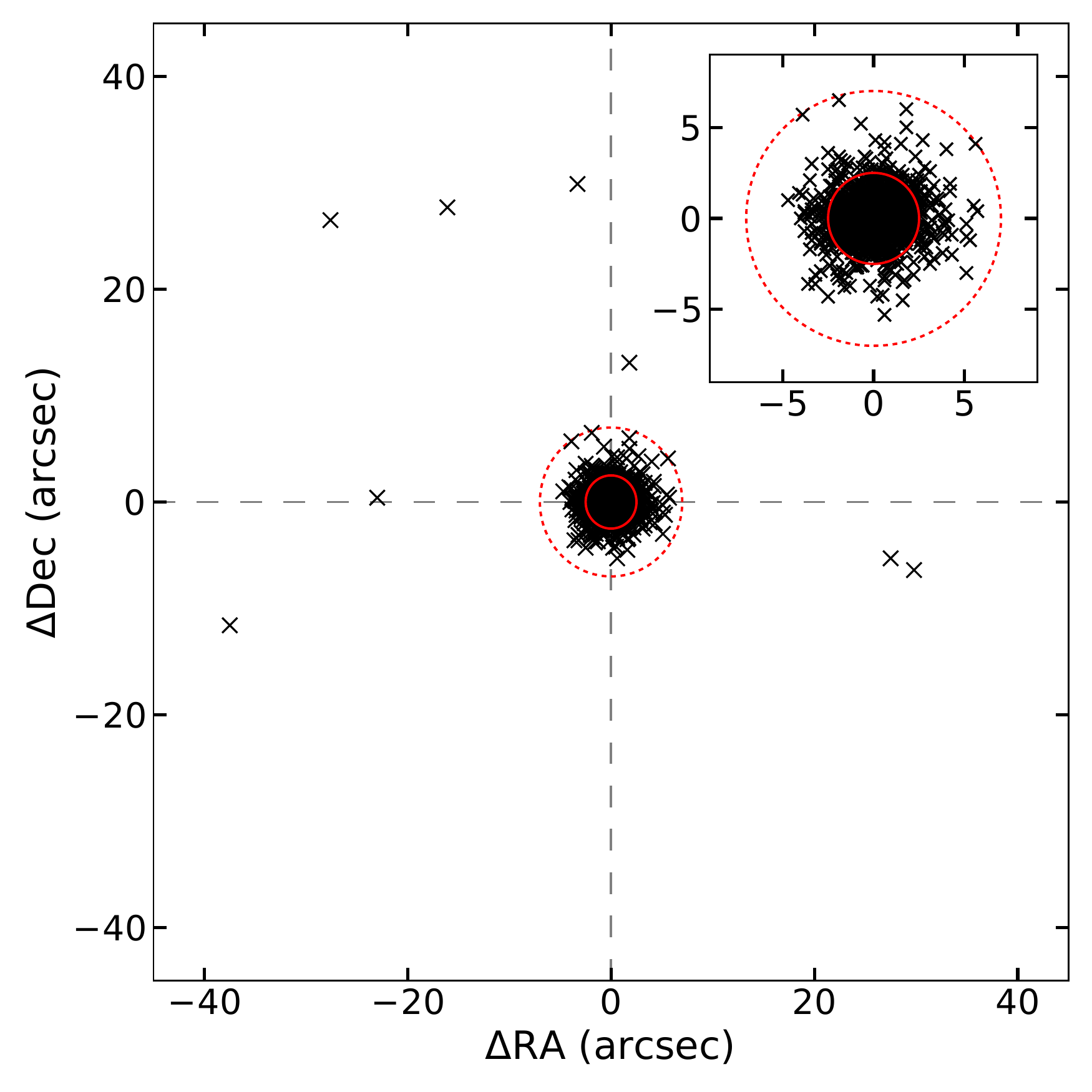}
\caption{Plot showing the offsets in position between the AT20G sources and their WISE counterparts. The inset shows the close-up of the central region of the image, with the solid circle drawn at 2.5 arcsec and dotted circle drawn at 7 arcsec radii. {The small number of sources with large offsets are due to large errors in their positions in the AT20G survey.}}
\label{Fig:posOffsets}
\end{figure}
\end{center}

A very small number of sources in \autoref{Fig:posOffsets} have large offsets ($>$10 arcsec) between the AT20G and WISE/optical positions. These were all checked by eye and in the majority of cases the offset is due to occasional large errors in the AT20G position when the wrong peak was chosen in a poor quality postage-stamp image in the original catalogue. 
All of these objects have an optical counterpart so these positions were used when making the WISE crossmatch, but if the AT20G positions were used these are likely to have been missed. This raises the possibility that there may be a small number of AT20G sources with inaccurate positions and no optical counterparts that are not included in this catalogue. Given the small number of sources that have a WISE match and no optical match we do not expect this to significantly effect the following analysis.

\begin{center}
\begin{figure}
\includegraphics[scale=0.39, angle=0]{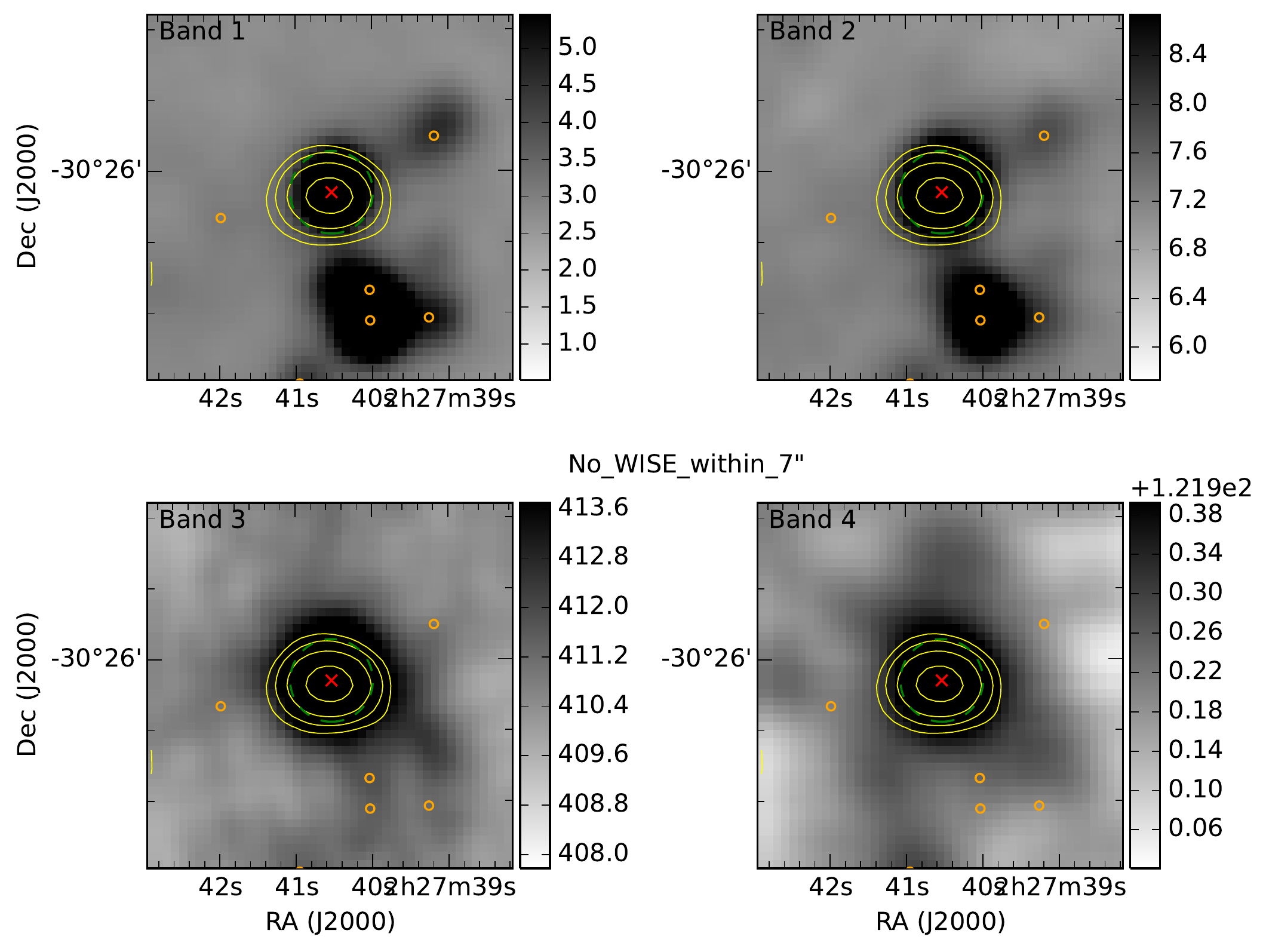}
\caption{Figure shows AT20G source J022740-302604 whose WISE counterpart is seen in images but not listed in the AllWISE catalogue. Photometry for this source was provided by T.~Jarrett (private communication).  The red cross is the search position, yellow contours are from the AT20G survey, the dotted green circle is the 7 arcsec search radius and the orange circles are identifications listed in the AllWISE catalogue.}
\label{Fig:NONE2}
\end{figure}
\end{center}

\section{The AT20G-WISE catalogue} \label{Sec:presentCatalogue}
We present the first fifty sources in the AT20G-WISE compact source catalogue in \autoref{Tab:main_table}. We have also included the AT20G-optical counterparts and their optical properties, based on \cite{Mahony2011}. The complete table of all sources is provided as additional data in the online version of the paper and will be available through the Vizier catalogue database.

\begin{landscape}
\begin{table}
\centering
\resizebox{\columnwidth}{!}
{
\begin{tabular}{cccrrrrlllccrrlllcccrrlllllllllllll}
\hline
AT20G& RA & DEC & $S_{20}$ & $S_{8.6}$ & $S_{4.8}$ & $\alpha_1^{4.8}$ & 6-km & Err & Flag & Op RA & Op Dec & B & R  & z & Op & Match & WISE & RA & Dec & RA & Dec & W1 & W1 & W2 & W2 & W3 & W3 & W4 & W4 & Mag & Quality & cc & Ph & Ext \\

Name & (J2000) & (J2000) & (mJy) & (mJy) & (mJy) &  & Vis &  &  & (J2000) & (J2000) & mag & mag &  & Flag & Flag & designation & (J2000) & (J2000) & offset ('') & offset ('') & mag & $\sigma$ & mag & $\sigma$ & mag & $\sigma$ & mag & $\sigma$ & Type & Flag & Flag & qual & Flag \\
(1) & (2) & (3) & (4) & (5) & (6) & (7) & (8) & (9) & (10) & (11) & (12) & (13) & (14) & (15) & (16) & (17) & (18) & (19) & (20) & (21) & (22) & (23) & (24) & (25) & (26) & (27) & (28) & (29) & (30) & (31) & (32) & (33) & (34) & (35) \\
\hline

J000012-853919 & 00:00:12.78 & -85:39:19.9 & 98 & 63.0 & 63.0 & -0.29 & 0.94 & 0.04 & v & 00:00:13.64 & -85:39:21.4 & 19.9 & . & . & b & Aa & J000013.20-853921.2 & 00:00:13.21 & -85:39:21.2 & 0.5 & -1.3 & 15.895 & 0.041 & 15.089 & 0.055 & 12.182 & 0.224 & 8.775 & . & pppp & 0001c & 0000c & AABU & 0\\
J000020-322101 & 00:00:20.38 & -32:21:01.2 & 118 & 315.0 & 515.0 & -0.01 & 0.97 & 0.04 & . & 00:00:20.38 & -32:21:01.2 & 18.7 & 18.3 & 1.275 & . & Aa & J000020.40-322101.2 & 00:00:20.40 & -32:21:01.2 & 0.3 & -0.0 & 14.812 & 0.032 & 13.458 & 0.031 & 10.198 & 0.068 & 7.731 & 0.168 & pppp & 0000 & 0000 & AAAB & 0\\
J000105-155107 & 00:01:05.42 & -15:51:07.2 & 297 & 295.0 & 257.0 & -0.24 & 0.93 & 0.03 & . & 00:01:05.33 & -15:51:07.0 & 18.2 & 18.4 & 2.044 & . & Aa & J000105.29-155107.2 & 00:01:05.29 & -15:51:07.2 & -1.9 & -0.0 & 15.12 & 0.036 & 13.938 & 0.04 & 10.511 & 0.09 & 8.371 & 0.383 & pppp & 0002 & 0000 & AAAC & 0\\
J000118-074626 & 00:01:18.04 & -07:46:26.8 & 177 & . & . & . & 0.91 & 0.03 & . & 00:01:18.02 & -07:46:26.9 & 18.0 & 17.1 & . & . & Aa & J000118.01-074626.9 & 00:01:18.02 & -07:46:26.9 & -0.4 & -0.1 & 12.702 & 0.024 & 11.752 & 0.023 & 9.163 & 0.036 & 6.975 & 0.111 & pppp & 0000 & 0000 & AAAB & 0\\
J000125-065624 & 00:01:25.59 & -06:56:24.7 & 77 & . & . & . & 0.93 & 0.05 & . & 00:01:25.58 & -06:56:24.9 & 19.3 & 18.6 & . & . & Aa & J000125.57-065625.0 & 00:01:25.58 & -06:56:25.1 & -0.2 & -0.4 & 15.524 & 0.044 & 14.212 & 0.054 & 11.409 & 0.225 & 8.443 & 0.345 & pppp & 0000 & 0000 & AABB & 0\\
J000249-211419 & 00:02:49.85 & -21:14:19.2 & 100 & . & . & . & 0.87 & 0.04 & . & 00:02:49.75 & -21:14:20.3 & 19.7 & 19.2 & . & . & Aa & J000249.75-211420.1 & 00:02:49.76 & -21:14:20.2 & -1.3 & -1.0 & 16.341 & 0.068 & 15.14 & 0.082 & 12.399 & 0.454 & 8.816 & . & pppp & 0221 & 0h00 & AACU & 0\\
J000252-594814 & 00:02:52.93 & -59:48:14.0 & 71 & 64.0 & 57.0 & -0.05 & 0.99 & 0.06 & . & 00:02:52.96 & -59:48:14.6 & . & 21.6 & . & . & Aa & J000252.91-594814.7 & 00:02:52.91 & -59:48:14.8 & -0.1 & -0.8 & 14.721 & 0.028 & 13.738 & 0.029 & 11.623 & 0.181 & 8.602 & . & pppp & 0001 & 0000 & AABU & 0\\
J000303-553007 & 00:03:03.45 & -55:30:07.1 & 44 & 48.0 & 52.0 & 0.09 & 1.05 & 0.09 & . & 00:03:03.33 & -55:30:07.2 & 22.2 & 20.7 & . & . & Aa & J000303.36-553007.1 & 00:03:03.36 & -55:30:07.2 & -0.7 & -0.1 & 16.644 & 0.076 & 15.449 & 0.088 & 11.754 & 0.225 & 8.722 & 0.376 & pppp & 0002 & 0000 & AABC & 0\\
J000316-194150 & 00:03:16.06 & -19:41:50.7 & 76 & 162.0 & 182.0 & -0.2 & 0.95 & 0.05 & . & 00:03:15.94 & -19:41:50.3 & 19.3 & 18.8 & . & . & Aa & J000315.94-194150.1 & 00:03:15.95 & -19:41:50.1 & -1.6 & 0.6 & 15.214 & 0.038 & 14.576 & 0.062 & 12.038 & 0.322 & 8.658 & . & pppp & 0001 & 0000 & AABU & 0\\
J000404-114858 & 00:04:04.88 & -11:48:58.0 & 680 & . & . & . & 1.01 & 0.03 & . & 00:04:04.90 & -11:48:58.4 & 18.1 & 18.6 & . & . & Aa & J000404.91-114858.3 & 00:04:04.91 & -11:48:58.3 & 0.5 & -0.3 & 14.258 & 0.028 & 13.217 & 0.031 & 10.349 & 0.076 & 8.479 & 0.348 & pppp & 0000 & 0000 & AAAB & 0\\
J000407-434510 & 00:04:07.24 & -43:45:10.0 & 199 & 211.0 & 244.0 & -0.2 & 0.88 & 0.03 & . & 00:04:07.24 & -43:45:10.0 & 18.5 & 18.3 & . & . & Aa & J000407.26-434510.1 & 00:04:07.26 & -43:45:10.2 & 0.2 & -0.2 & 14.407 & 0.029 & 13.168 & 0.029 & 10.181 & 0.063 & 8.228 & 0.333 & pppp & 0000 & 0000 & AAAB & 0\\
J000435-473619 & 00:04:35.65 & -47:36:19.0 & 868 & 970.0 & 900.0 & -0.02 & 0.98 & 0.03 & . & 00:04:35.64 & -47:36:19.5 & 17.6 & 17.6 & 0.884 & . & Aa & J000435.65-473619.6 & 00:04:35.66 & -47:36:19.7 & 0.1 & -0.7 & 13.228 & 0.009 & 12.173 & 0.013 & 8.78 & 0.026 & 6.474 & 0.066 & 44pp & 2222 & 0000 & AAAA & 1\\
J000507-013244 & 00:05:07.03 & -01:32:44.6 & 81 & . & . & . & 0.96 & 0.05 & . & 00:05:07.06 & -01:32:45.2 & 18.7 & 18.3 & 1.71 & . & Aa & J000507.07-013245.2 & 00:05:07.07 & -01:32:45.3 & 0.7 & -0.7 & 15.785 & 0.055 & 14.829 & 0.075 & 11.808 & 0.338 & 8.631 & 0.44 & pppp & 0002 & 0000 & AABC & 0\\
J000518-164804 & 00:05:18.01 & -16:48:04.9 & 142 & . & . & . & 0.91 & 0.04 & . & 00:05:17.93 & -16:48:04.7 & 18.0 & 17.7 & 0.777 & . & Aa & J000517.92-164804.4 & 00:05:17.93 & -16:48:04.5 & -1.2 & 0.4 & 13.643 & 0.026 & 12.619 & 0.026 & 10.158 & 0.057 & 7.745 & 0.166 & pppp & 0000 & 0000 & AAAB & 0\\
J000600-313215 & 00:06:00.47 & -31:32:15.0 & 63 & 53.0 & 52.0 & 0.13 & 0.91 & 0.06 & . & 00:06:00.37 & -31:32:14.8 & 21.1 & 19.9 & . & . & Aa & J000600.36-313214.8 & 00:06:00.37 & -31:32:14.9 & -1.3 & 0.1 & 15.067 & 0.035 & 14.27 & 0.044 & 11.87 & 0.257 & 8.601 & . & pppp & 0001 & 0000 & AABU & 0\\
J000601-295549 & 00:06:01.14 & -29:55:49.6 & 97 & 187.0 & 228.0 & 0.78 & 0.86 & 0.05 & . & 00:06:01.12 & -29:55:50.0 & 18.1 & 18.8 & . & . & Aa & J000601.12-295550.0 & 00:06:01.12 & -29:55:50.0 & -0.2 & -0.4 & 13.735 & 0.025 & 12.499 & 0.024 & 10.129 & 0.061 & 7.952 & 0.242 & pppp & 0000 & 0000 & AAAB & 0\\
J000613-062334 & 00:06:13.90 & -06:23:34.8 & 2084 & . & . & . & 0.99 & 0.03 & . & 00:06:13.89 & -06:23:35.3 & 19.5 & 18.0 & 0.347 & . & Aa & J000613.88-062335.2 & 00:06:13.89 & -06:23:35.3 & -0.2 & -0.5 & 12.151 & 0.023 & 11.009 & 0.02 & 8.022 & 0.021 & 5.52 & 0.042 & pppp & 0000 & 0000 & AAAA & 0\\
J000619-424518 & 00:06:19.72 & -42:45:18.3 & 183 & 189.0 & 199.0 & -0.12 & 0.92 & 0.03 & . & 00:06:19.72 & -42:45:18.5 & 18.3 & 18.4 & 1.77 & . & Aa & J000619.73-424518.6 & 00:06:19.73 & -42:45:18.6 & 0.2 & -0.3 & 15.449 & 0.04 & 14.331 & 0.044 & 11.202 & 0.126 & 8.372 & 0.253 & pppp & 0000 & 0000 & AABB & 0\\
J000635-731144 & 00:06:35.21 & -73:11:44.7 & 51 & 54.0 & 54.0 & -0.09 & 1.02 & 0.08 & . & . & . & . & . & . & . & Aa & J000635.43-731144.6 & 00:06:35.43 & -73:11:44.7 & 1.0 & 0.0 & 16.702 & 0.068 & 15.876 & 0.109 & 12.984 & 0.532 & 8.515 & . & pppp & 0021 & 0000 & AACU & 0\\
J000713-402337 & 00:07:13.41 & -40:23:37.4 & 69 & 89.0 & 97.0 & -0.01 & 0.9 & 0.06 & . & 00:07:13.42 & -40:23:37.4 & . & 19.8 & . & . & Aa & J000713.40-402337.1 & 00:07:13.40 & -40:23:37.1 & -0.1 & 0.3 & 16.811 & 0.084 & 15.987 & 0.146 & 12.644 & 0.494 & 8.93 & . & pppp & 0021 & 0000 & ABCU & 0\\
J000720-611306 & 00:07:20.56 & -61:13:06.7 & 150 & 138.0 & 134.0 & 0.26 & 0.97 & 0.04 & v & 00:07:20.55 & -61:13:06.7 & 19.9 & 20.8 & 0.857 & . & Aa & J000720.56-611306.7 & 00:07:20.57 & -61:13:06.8 & 0.1 & -0.1 & 16.074 & 0.049 & 15.119 & 0.065 & 12.177 & 0.284 & 8.84 & . & pppp & 0001 & 0000 & AABU & 0\\
J000800-233918 & 00:08:00.42 & -23:39:18.0 & 154 & . & . & . & 0.94 & 0.03 & . & 00:08:00.36 & -23:39:18.3 & 16.8 & 16.4 & 1.411 & . & Aa & J000800.37-233918.0 & 00:08:00.37 & -23:39:18.1 & -0.6 & -0.1 & 13.845 & 0.026 & 12.466 & 0.025 & 9.144 & 0.032 & 6.779 & 0.097 & pppp & 0000 & 0000 & AAAA & 0\\
J000801-524339 & 00:08:01.71 & -52:43:39.9 & 124 & 127.0 & 101.0 & -0.53 & 0.98 & 0.04 & . & 00:08:01.67 & -52:43:39.8 & 18.9 & 19.5 & . & . & Aa & J000801.67-524340.0 & 00:08:01.67 & -52:43:40.0 & -0.3 & -0.1 & 14.903 & 0.031 & 13.985 & 0.037 & 10.94 & 0.1 & 7.883 & 0.236 & pppp & 0000 & 0000 & AAAB & 0\\
J000826-255911 & 00:08:26.27 & -25:59:11.2 & 120 & 259.0 & 360.0 & -0.21 & 0.92 & 0.04 & . & 00:08:26.10 & -25:59:11.3 & . & 20.0 & . & . & Nv & . & . & . & . & . & . & . & . & . & . & . & . & . & . & . & . & . & . \\
J000828-132930 & 00:08:28.02 & -13:29:30.3 & 161 & . & . & . & 0.96 & 0.03 & . & . & . & . & . & . & . & Nv & . & . & . & . & . & . & . & . & . & . & . & . & . & . & . & . & . & . \\
J000837-461940 & 00:08:37.53 & -46:19:40.8 & 75 & 98.0 & 95.0 & -0.1 & 1.0 & 0.05 & . & 00:08:37.54 & -46:19:40.7 & 17.4 & 16.6 & 1.85 & . & Aa & J000837.54-461940.8 & 00:08:37.55 & -46:19:40.9 & 0.2 & -0.1 & 13.509 & 0.026 & 12.51 & 0.023 & 9.257 & 0.029 & 6.995 & 0.079 & pppp & 0000 & 0000 & AAAA & 0\\
J000922-513011 & 00:09:22.06 & -51:30:11.9 & 49 & 59.0 & 56.0 & 0.12 & 1.11 & 0.08 & . & 00:09:22.03 & -51:30:11.6 & 16.8 & 15.6 & 0.117 & . & As & J000922.01-513011.4 & 00:09:22.02 & -51:30:11.5 & -0.4 & 0.4 & 12.816 & 0.024 & 12.623 & 0.026 & 11.158 & 0.116 & 8.942 & 0.395 & mmmm & 0000 & . & . & 1\\
J000927-161028 & 00:09:27.55 & -16:10:28.8 & 69 & 92.0 & 61.0 & 0.08 & 0.95 & 0.06 & . & 00:09:27.50 & -16:10:29.2 & 20.9 & 20.6 & . & . & Rb & . & . & . & . & . & . & . & . & . & . & . & . & . & . & . & . & . & . \\
J000937-635734 & 00:09:37.40 & -63:57:34.2 & 59 & 66.0 & 75.0 & -0.18 & 0.98 & 0.07 & . & 00:09:37.51 & -63:57:33.8 & 20.5 & . & . & . & Aa & J000937.48-635733.7 & 00:09:37.49 & -63:57:33.8 & 0.6 & 0.4 & 15.828 & 0.044 & 14.761 & 0.05 & 12.074 & 0.254 & 8.899 & . & pppp & 0001 & 0000 & AABU & 0\\
J001000-043348 & 00:10:00.41 & -04:33:48.2 & 275 & . & . & . & 0.99 & 0.03 & . & 00:10:00.37 & -04:33:48.4 & 21.2 & 20.0 & . & . & Aa & J001000.38-043348.4 & 00:10:00.38 & -04:33:48.5 & -0.4 & -0.3 & 14.618 & 0.033 & 13.662 & 0.038 & 11.269 & 0.186 & 8.472 & . & pppp & 0001 & 0000 & AABU & 0\\
J001026-413001 & 00:10:26.05 & -41:30:01.8 & 63 & 65.0 & 71.0 & -0.36 & 1.05 & 0.06 & . & 00:10:26.03 & -41:30:02.1 & 18.8 & 18.4 & . & . & Aa & J001026.05-413002.1 & 00:10:26.05 & -41:30:02.2 & 0.1 & -0.4 & 14.597 & 0.029 & 13.612 & 0.032 & 10.704 & 0.094 & 7.812 & 0.172 & pppp & 0000 & 0000 & AAAB & 0\\
J001035-302748 & 00:10:35.92 & -30:27:48.2 & 741 & 718.0 & 629.0 & 0.56 & 0.96 & 0.03 & . & 00:10:35.72 & -30:27:47.4 & 19.6 & 18.7 & 1.19 & o & Aa & J001035.72-302747.5 & 00:10:35.73 & -30:27:47.5 & -2.5 & 0.7 & 14.795 & 0.032 & 13.778 & 0.036 & 10.748 & 0.099 & 8.059 & 0.196 & pppp & 0000 & 0000 & AAAB & 0\\
J001045-294513 & 00:10:45.16 & -29:45:13.0 & 118 & 335.0 & 503.0 & 0.18 & 0.92 & 0.04 & . & 00:10:45.16 & -29:45:13.1 & 18.6 & 18.5 & 2.353 & . & Aa & J001045.17-294513.4 & 00:10:45.17 & -29:45:13.5 & 0.2 & -0.5 & 16.066 & 0.054 & 14.849 & 0.066 & 11.219 & 0.177 & 8.673 & 0.401 & pppp & 0002 & 0000 & AABC & 0\\
J001052-415310 & 00:10:52.51 & -41:53:10.2 & 152 & 612.0 & 1351.0 & -0.89 & 0.9 & 0.04 & . & . & . & . & . & . & . & Aa & J001052.52-415310.1 & 00:10:52.52 & -41:53:10.2 & 0.1 & 0.0 & 16.678 & 0.076 & 15.905 & 0.133 & 11.68 & 0.224 & 8.521 & 0.293 & pppp & 0000 & 0000 & ABBB & 0\\
J001053-215704 & 00:10:53.69 & -21:57:04.0 & 367 & . & . & . & 0.92 & 0.03 & . & 00:10:53.64 & -21:57:03.8 & 21.9 & 20.6 & . & . & Aa & J001053.64-215704.2 & 00:10:53.64 & -21:57:04.3 & -0.6 & -0.3 & 14.801 & 0.032 & 13.786 & 0.037 & 10.566 & 0.08 & 7.699 & 0.191 & pppp & 2200 & hh00 & AAAB & 0\\
J001055-274224 & 00:10:55.29 & -27:42:24.6 & 84 & 88.0 & 54.0 & -0.08 & 0.95 & 0.05 & . & 00:10:55.26 & -27:42:25.0 & 19.2 & 18.7 & 1.45 & . & Aa & J001055.28-274225.6 & 00:10:55.28 & -27:42:25.6 & -0.1 & -1.0 & 15.36 & 0.043 & 14.208 & 0.047 & 11.104 & 0.133 & 8.819 & 0.405 & pppp & 0002 & 0000 & AABC & 0\\
J001101-261233 & 00:11:01.27 & -26:12:33.1 & 642 & 822.0 & 694.0 & 0.97 & 0.96 & 0.03 & . & 00:11:01.21 & -26:12:33.1 & 19.5 & 18.9 & 1.096 & . & Aa & J001101.22-261232.0 & 00:11:01.23 & -26:12:32.0 & -0.5 & 1.1 & 14.719 & 0.031 & 14.024 & 0.042 & 10.909 & 0.122 & 8.692 & 0.457 & pppp & 0002 & 0000 & AABC & 0\\
J001140-143404 & 00:11:40.44 & -14:34:04.2 & 154 & . & . & . & 0.94 & 0.03 & . & . & . & . & . & . & . & Rb & . & . & . & . & . & . & . & . & . & . & . & . & . & . & . & . & . & . \\
J001146-844319 & 00:11:46.28 & -84:43:19.4 & 274 & 260.0 & 236.0 & 0.02 & 0.97 & 0.03 & v & 00:11:45.87 & -84:43:20.0 & 18.9 & 18.9 & . & . & Aa & J001145.90-844320.0 & 00:11:45.90 & -84:43:20.0 & -0.5 & -0.6 & 14.665 & 0.028 & 13.517 & 0.029 & 10.721 & 0.066 & 8.17 & 0.165 & pppp & 0000 & 0000 & AAAB & 0\\
J001150-870625 & 00:11:50.52 & -87:06:25.0 & 57 & 110.0 & 137.0 & 0.1 & 1.11 & 0.07 & . & 00:11:51.38 & -87:06:25.4 & 18.1 & 17.5 & . & . & Aa & J001151.57-870625.5 & 00:11:51.58 & -87:06:25.5 & 0.8 & -0.5 & 14.105 & 0.026 & 13.654 & 0.03 & 10.808 & 0.075 & 8.242 & 0.239 & pppp & 0000 & 0000 & AAAB & 0\\
J001152-410545 & 00:11:52.38 & -41:05:45.1 & 132 & 163.0 & 175.0 & -0.18 & 0.96 & 0.04 & . & 00:11:52.38 & -41:05:45.2 & 20.6 & 18.6 & . & . & Aa & J001152.39-410545.1 & 00:11:52.40 & -41:05:45.2 & 0.2 & -0.1 & 15.279 & 0.035 & 14.255 & 0.04 & 11.361 & 0.116 & 8.92 & 0.357 & pppp & 0002 & 0000 & AABC & 0\\
J001259-395426 & 00:12:59.89 & -39:54:26.4 & 1609 & 2009.0 & 2015.0 & 1.14 & 0.96 & 0.03 & . & 00:12:59.89 & -39:54:25.9 & 18.3 & 18.0 & . & . & Aa & J001259.89-395425.9 & 00:12:59.90 & -39:54:25.9 & 0.1 & 0.5 & 13.809 & 0.024 & 12.776 & 0.024 & 9.894 & 0.042 & 7.539 & 0.101 & pppp & 0000 & 0000 & AAAA & 0\\
J001320-151347 & 00:13:20.72 & -15:13:47.9 & 189 & . & . & . & 0.92 & 0.03 & . & 00:13:20.70 & -15:13:47.7 & 19.3 & 19.0 & 1.838 & . & Aa & J001320.71-151347.6 & 00:13:20.71 & -15:13:47.6 & -0.1 & 0.3 & 16.244 & 0.062 & 15.071 & 0.084 & 11.993 & 0.322 & 9.083 & . & pppp & 0001 & 0000 & AABU & 0\\
J001324-500514 & 00:13:24.21 & -50:05:14.1 & 71 & 58.0 & 57.0 & 0.1 & 1.01 & 0.06 & . & 00:13:24.18 & -50:05:14.4 & 19.5 & 18.5 & . & . & Aa & J001324.18-500514.5 & 00:13:24.19 & -50:05:14.6 & -0.2 & -0.5 & 15.247 & 0.035 & 14.093 & 0.038 & 11.452 & 0.158 & 8.24 & . & pppp & 0001 & 0000 & AABU & 0\\
J001354-042352 & 00:13:54.14 & -04:23:52.0 & 550 & . & . & . & 0.98 & 0.03 & . & 00:13:54.12 & -04:23:52.1 & 20.2 & 19.7 & 1.075 & . & Aa & J001354.13-042352.1 & 00:13:54.13 & -04:23:52.2 & -0.1 & -0.2 & 15.652 & 0.051 & 14.636 & 0.07 & 11.057 & 0.156 & 8.465 & 0.414 & pppp & 0002 & 0000 & AABC & 0\\
J001427-563822 & 00:14:27.01 & -56:38:22.6 & 134 & 163.0 & 179.0 & 0.42 & 0.97 & 0.04 & . & . & . & . & . & . & . & Aa & J001426.83-563822.8 & 00:14:26.84 & -56:38:22.8 & -1.4 & -0.2 & 16.039 & 0.052 & 15.029 & 0.064 & 11.355 & 0.162 & 9.034 & 0.507 & pppp & 0002 & 0000 & AABC & 0\\
J001434-200031 & 00:14:34.53 & -20:00:31.9 & 118 & . & . & . & 0.94 & 0.04 & . & 00:14:34.46 & -20:00:31.7 & 18.8 & 18.2 & . & . & Aa & J001434.46-200031.8 & 00:14:34.46 & -20:00:31.8 & -1.0 & 0.1 & 14.464 & 0.029 & 13.433 & 0.032 & 11.098 & 0.147 & 8.261 & 0.289 & pppp & 0000 & 0000 & AABB & 0\\
J001502-181250 & 00:15:02.53 & -18:12:50.6 & 355 & . & . & . & 0.96 & 0.03 & . & 00:15:02.49 & -18:12:51.0 & 19.7 & 19.0 & 0.743 & . & Aa & J001502.49-181250.8 & 00:15:02.50 & -18:12:50.9 & -0.5 & -0.3 & 14.828 & 0.031 & 13.625 & 0.034 & 11.403 & 0.223 & 8.503 & 0.343 & pppp & 0000 & 0000 & AABB & 0\\
J001534-180725 & 00:15:34.39 & -18:07:25.4 & 98 & . & . & . & 0.97 & 0.04 & . & . & . & . & . & . & . & Rv & . & . & . & . & . & . & . & . & . & . & . & . & . & . & . & . & . & . \\
J001552-451103 & 00:15:52.91 & -45:11:03.3 & 70 & 74.0 & 80.0 & -0.36 & 0.93 & 0.06 & . & 00:15:52.94 & -45:11:03.4 & 22.4 & 21.0 & . & . & Aa & J001552.94-451103.2 & 00:15:52.95 & -45:11:03.3 & 0.4 & 0.0 & 15.492 & 0.041 & 14.76 & 0.065 & 12.234 & 0.432 & 8.509 & . & pppp & 0021 & 0000 & AACU & 0\\

\hline\\
\end{tabular}%
}
\medskip\\
\caption{We present the first 50 sources in the catalogue cross-matching the AT20G compact AGNs with the AllWISE mid-infrared catalogue. Empty fields have been replaced with a `.' sign to aid with visual inspection in this table only. Uncertainty in flux densities for AT20G are provided in the AT20G catalogue \citep{Murphy2010}. Using typical uncertainty in flux density for the AT20G survey ($\sim$5 per cent) and  for NVSS and SUMSS surveys ($\sim$ 4 per cent for both), uncertainty for spectral index in column 7 are typically $\sim$0.02. 
Full table is available as supplementary material to this paper and, via the \textit{VizieR} Catalogue Service.}
\label{Tab:main_table}
\end{table}
\end{landscape}

Descriptions of the different columns in the catalogue are provided below. For brevity, we have not reproduced uncertainty values for flux densities in columns 4, 5 and 6, which are to be found in \cite{Murphy2010}. Using typical uncertainty in flux density for the AT20G survey ($\sim$5 per cent) and  for NVSS \citep{Condon1998} and SUMSS \citep{Mauch2003} surveys ($\sim$ 4 per cent for both), we estimate uncertainty for spectral index in column 7 to be typically $\sim$0.02. The WISE catalogue provides multiple and complicated quality flags necessary to determine the quality of photometry. We have re-produced relevant ones in our catalogue which we briefly describe below. 
\begin{tabbing}
	$Col 1$ ~~~~~\= AT20G name for the source,\\
	$Col 2$   \> Right Ascension (J2000),\\
	$Col 3$   \> Declination (J2000),\\
	$Col 4$   \> 20 GHz flux density (mJy) from the AT20G survey, \\ 
	$Col 5$   \> 8.6 GHz flux density (mJy) from the AT20G survey,\\
	$Col 6$   \> 4.8 GHz flux density (mJy) from the AT20G survey,\\
	$Col 7$   \> Spectral index between 1 (either NVSS or SUMSS)\\\> and 4.8 GHz,\\
	$Col 8$   \> 6-km visibility,\\
	$Col 9$   \> error in the 6-km visibility,\\
	$Col 10$  \> flags from AT20G high-angular-resolution catalogue\\
		  \>\citep[]{Chhetri2013}. Sources can have more  \\\>than one flag:\\
		  \>\textit{b : source is large and extended in AT20G}\\\>~~~~\textit{main catalogue,}\\
		  \>\textit{e : source is flagged as extended in AT20G}\\\>~~~~\textit{main catalogue,}\\
		  \>\textit{l : source has no counterpart in either NVSS}\\\>~~~~\textit{or SUMSS,}\\
		  \>\textit{m : part of Magellanic clouds,}\\
		  \>\textit{v : more than two 6-km visibility values are present},\\
	$Col 11$  \> Right Ascension of optical counterpart (J2000),\\
	$Col 12$  \> Declination of optical counterpart (J2000),\\
	$Col 13$  \> B magnitude from SuperCOSMOS, \\
	$Col 14$  \> R magnitude from SuperCOSMOS, \\
	$Col 15$  \> redshift, where available, \\
	$Col 16$  \> optical flags from \cite{Mahony2011}:\\
		  \>\textit{b : blended source,}\\
		  \>\textit{c : source is in a field too crowded to make an} \\ \>\textit{~~~~identification,}\\
		  \>\textit{o : source is slightly offset from radio position} \\ \>\textit{~~~~but still within 3.5 arcsec,}\\
		  \>\textit{u : unreliable optical magnitude},\\
	$Col 17$  \> Match flag: \\
		  \>\textit{Aa : cross-match accepted automatically,}\\
		  \>\textit{Am : cross-match obtained manually,}\\
		  \>\textit{As : taken from \cite{Sadler2014},}\\
		  \>\textit{Av : cross-match accepted after visual inspection},\\
		  \>\textit{Nv : source has no counterpart},\\
		  \>\textit{Rb : cross-match rejected as counterpart is blended},\\
		  \>\textit{Rv : cross-match rejected after visual inspection},\\ 
	$Col 18$  \> WISE designation of accepted counterpart, common \\\>names of a small number of objects are presented in \\\>\autoref{Tab:commonNames}\\
	$Col 19$  \> Right Ascension of accepted counterpart (J2000), \\
	$Col 20$  \> Declination of accepted counterpart (J2000),\\
	$Col 21$  \> offset in RA in WISE counterpart position from\\\> AT20G position (arcsec), \\
	$Col 22$  \> offset in Dec in WISE counterpart position\\\>from AT20G position (arcsec), \\
	$Col 23$  \> WISE magnitude in band 1 (3.4\,micron),\\
	$Col 24$  \> instrumental profile-fit photometry flux\\ \>uncertainty in mag units in band 1, \\	
	$Col 25$  \> WISE magnitude in band 2 (4.6\,micron),\\
	$Col 26$  \> instrumental profile-fit photometry flux\\ \>uncertainty in mag units in band 2, \\	
	$Col 27$  \> WISE magnitude in band 3 (12\,micron),\\
	$Col 28$  \> instrumental profile-fit photometry flux\\ \>uncertainty in mag units in band 3, \\	
	$Col 29$  \> WISE magnitude in band 4 (22\,micron),\\
	$Col 30$  \> instrumental profile-fit photometry flux\\ \>uncertainty in mag units in band 4, \\	
	$Col 31$  \> magnitude type, each letter/number represents\\\> magnitude type for each band from left (W1)\\\>to right (W4),\\
		  \>\textit{1-5 : magnitude is obtained from aperture 1 to}\\ \> \textit{ aperture 5}\\
		  \>\textit{g   : gmag,}\\
		  \>\textit{m   : magnitude was obtained manually,}\\
		  \>\textit{p   : profile-fit magnitude,}\\
		  \>\textit{.   : no magnitude available},\\
	$Col 32$  \> quality flag for magnitude, each number represents\\\>magnitude quality for each band from left (W1)\\\>to right (W4),\\
		  \>\textit{0 : excellent detection,}\\
		  \>\textit{1 : source photometry is upper limit,}\\
		  \>\textit{2 : source photometry should be used with caution,}\\
		  \>\textit{3 : bad photometry, do not use,}\\
	$Col 33$  \> contamination and confusion flag from WISE\\ \>catalogue, each number represents cc flag for each\\\>band from left (W1) to right (W4),\\
		  \>\textit{D,d : affected by diffraction spike,}\\
		  \>\textit{P,p : affected by persistency, left by a bright image,}\\
		  \>\textit{H,h : affected by halo from a nearby brigth source,}\\
		  \>\textit{O,o : (letters ``O,o'') affected by optical ghost due}\\\>~~~~~~~\textit{to a nearby bright source,}\\
		  \>\textit{0~~ : (number zero), source unaffected by known}\\\>~~~~~~\textit{artifacts,}\\
	$Col 34$  \> photometric quality flag from WISE catalogue,\\ \>each letter represents ph\_qual for each band from\\ \>left (W1) to right (W4),\\
		  \>\textit{A : W?SNR$\geq$10, where ? = bands 1,2,3 or 4,}\\
		  \>\textit{B : 3$<$W?SNR$<$10,}\\
		  \>\textit{C : 2$<$W?SNR$<$3,}\\
		  \>\textit{U : upper limit on magnitude,}\\
		  \>\textit{X : profile-fit measurement not possible,}\\
	$Col 35$  \> extended source flag from WISE catalogue (see \\ \>text for further explanation),\\
		  \>\textit{0 : not extended,}\\
		  \>\textit{1 : w?chi2$\geq$3.0 in any band, suspected to be}\\ \>~~~~~\textit{resolved or confused,}\\
		  \>\textit{2 : within isophotal extent of a 2MASS extended}\\\>~~~~~\textit{source,}\\
		  \>\textit{3 : as above and w?chi2$\geq$3.0 in any band,}\\
		  \>\textit{4 : identically associated with a 2MASS extended}\\\>~~~~~\textit{source,}\\
		  \>\textit{5 : as above and also w?chi2$\geq$3.0 in any band.}
	\end{tabbing}

Further detail about the flags for WISE cross-matches in \autoref{Tab:main_table} are provided in Appendix~\ref{Appendix:Flags}. Notes on special cases for highly extended sources are provided in Appendix \ref{Appendix:HighlyExtendeSources} and \ref{Appendix:CenA}.

\subsection{Use of appropriate WISE magnitude}
The WISE catalogues provide several definitions of magnitude (based on aperture and profile fitting). The usage of the correct magnitude definition depends upon the observed profile of the source, i.e.~if the source is unresolved or has an extended profile. Approximately 87 per cent of our sources are unresolved in WISE, therefore, we used W?MPRO (where ? is the observing band 1, 2, 3 or 4), which is obtained using a point source profile fit. For sources that are identified with or superimposed on known galaxies or have been otherwise ascertained to be extended, the aperture magnitude (W?MAG) gives a better estimate of the source magnitude.  For sources that are in the 2MASS Extended Source Catalogue (XSC), a scaled elliptical aperture matching the 2MASS XSC source is defined to obtain the ``g?mag'' and gives the best estimate of the source magnitude. Please find further details on extended flag (ext$\_$flg) in Appendix \ref{Appendix:Flags}.

\section{Discussion}\label{sec:Discussion}
\subsection{Radio AGNs in WISE colour-colour plots}
\begin{center}
\begin{figure}
\includegraphics[scale=0.55, angle=0]{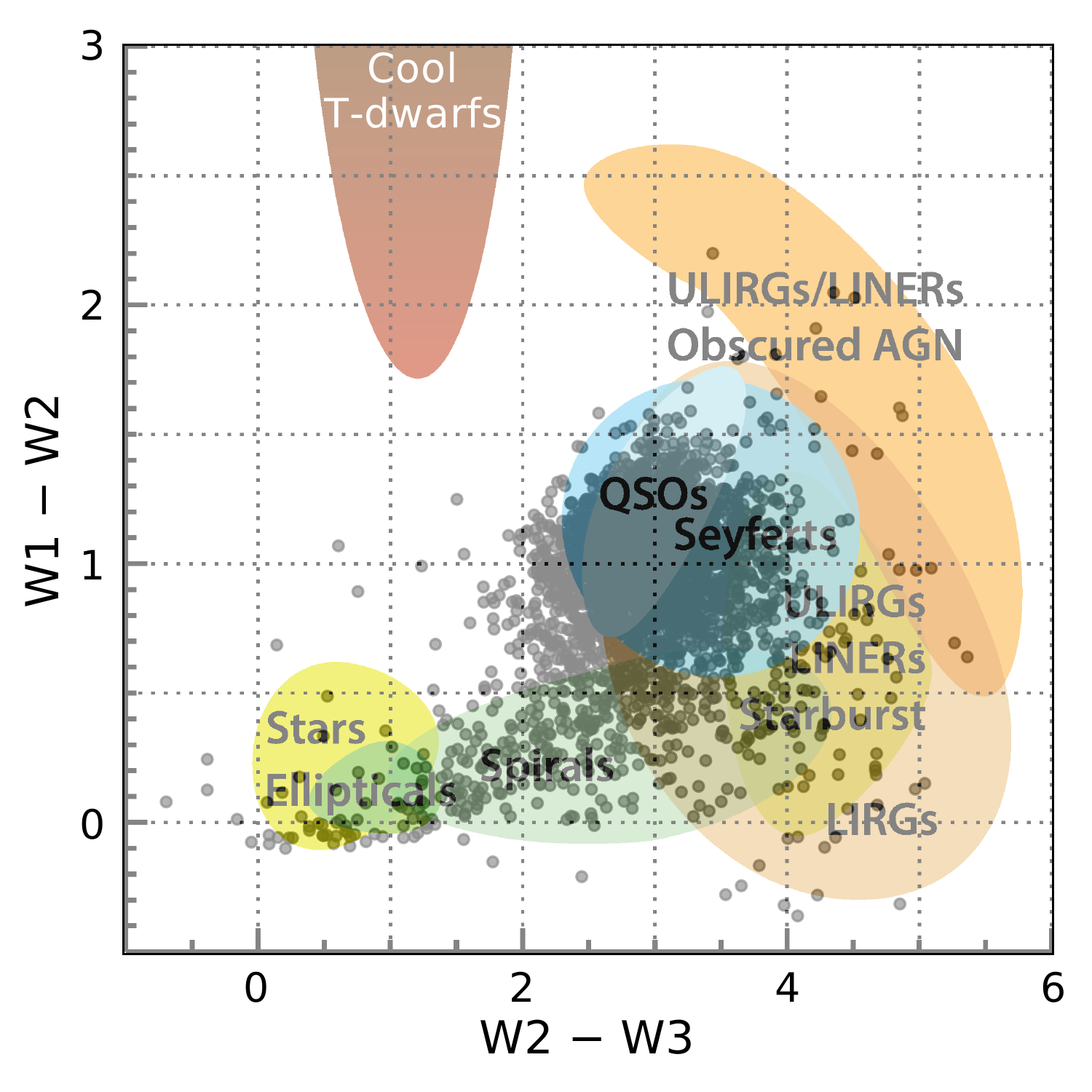}
\caption{The distribution of AT20G compact AGNs (black filled circles) in the WISE colour-colour plot, with figure 12 from \protect\cite{Wright2010} in the background.} 
\label{Fig:color-color_overlaid}
\end{figure}
\end{center}

\cite{Wright2010} have used the WISE colours of different source populations in the first three WISE bands to produce a ``template'' that outlines positions occupied by these source populations in the WISE colour-colour plot (their figure 12). In \autoref{Fig:color-color_overlaid}, we present a similar plot for the AT20G AGNs using their WISE colours presented in our catalogue. The figure shows evidence that the radio selected compact AGNs are a mixture of different populations as classified from mid-infrared colours, primarily dominated by QSOs and Seyfert galaxies, and a sizeable normal galaxy population.

A small number of sources with W2-W3$<$2 appear to have high W1-W2 values. All these objects have a quality flag of 2 in all four bands indicating that their magnitudes should be used with caution. Thus, these are likely to be normal galaxies whose W1-W2 colours are due to lower quality measurement. We further note that for a small number of cases (approximately 3 per cent) false identifications can cause sources to appear as outliers in this diagram. Barring these, outliers in the different corners of the plot with good quality measurements are genuine outliers with interesting properties.

We note that the classifications of QSOs, Seyferts and normal galaxies used here (and in subsequent sections) are based on WISE colours alone, and that these classifications may suffer from effects such as host galaxy contamination and other biases (such that, e.g., weak AGNs may not be detectable in mid-infrared). 

\subsection{Optical identifications and redshift distribution}\label{sec:OpticalCounterparts}
Optical identifications of our sources are taken from \cite{Mahony2011}, who used the digitized UK Schmidt Telescope survey in the SuperCOSMOS database; redshift information was obtained from the 6dF galaxy survey, new observations or the literature. We find 2935 objects have optical identifications and 1051 have redshift information.

\begin{center}
\begin{figure}
\includegraphics[scale=0.4, angle=0]{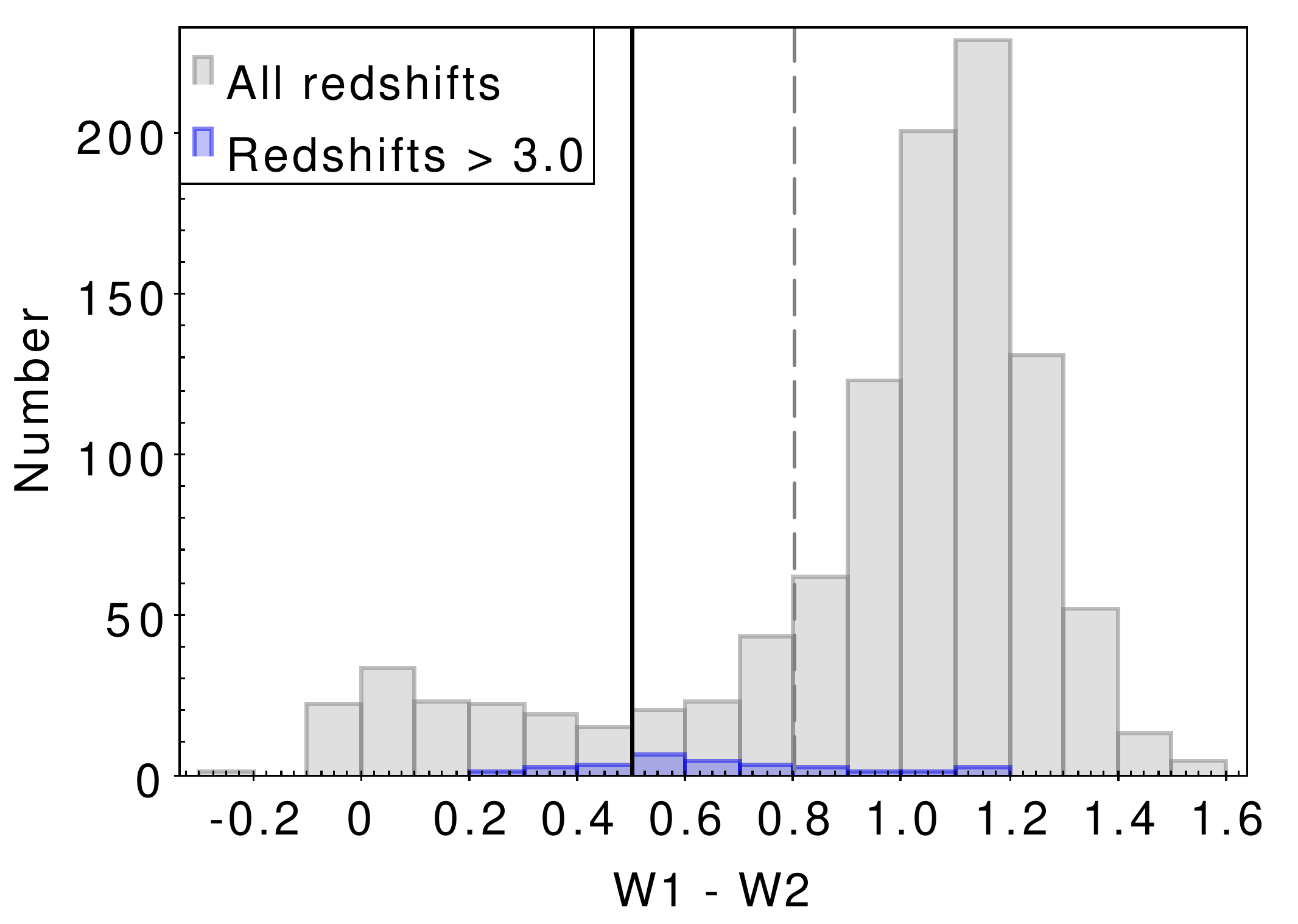}
\caption{Histogram of W1$-$W2 colours for all sources with redshift (light grey). Histogram for sources with redshift $>$ 3.0 are shown with dark bars. The solid line is drawn at W1-W2=0.5, where the AGN and galaxy population separate, a similar cut-off has also been used by \protect\cite{Satyapal2014ApJ...784..113S}, while the dashed line shows the cut-off used by \protect\cite{Stern2012}.}
\label{Fig:histo-z_W1W2}
\end{figure}
\end{center}

In the literature, the WISE W1$-$W2 colour has often been used to identify pure AGNs. For example \cite{Stern2012} have used a colour selection in WISE of W1-W2 $\geq$ 0.8 to identify pure AGNs up to redshift of $\sim$3, based on an AGN SED template by \cite{Assef2010}. 
Other authors \citep{Jarrett2011, Mateos2012} have defined wedges in the WISE colour-colour plot (such as \autoref{Fig:color-color_overlaid}) to idenfity AGNs. \cite{Hickox2017} conclude that luminous quasars can be effectively selected using a simple criteria such as those identified by \cite{Stern2012}, although they warn that such criteria may miss heavily obscured quasars.
In \autoref{Fig:histo-z_W1W2}, we plot the histogram of W1-W2 colours of all sources with redshift in our sample. For our data, we find that the limit of W1$-$W2~=~0.5 can be used to efficiently separate AGNs from galaxies, correctly classifying approximately 50 per cent more objects with redshift $>$3.0 as AGNs compared to when using the cut-off of 0.8 (see also \autoref{Fig:compareAssef}). 
Such a cut at W1$-$W2~=~0.5 has also been employed by \cite{Satyapal2014ApJ...784..113S} to identify bulgeless galaxies where the AGN emission dominates over the host galaxy emission.
While we find that this limit in W1$-$W2 has high efficiency to select AGNs in the case of strong compact sources selected at high radio frequencies, the removal of such radio pre-selection may render the cut-off less effective. For example, star-forming dwarf galaxies are also found in the region defined by W1$-$W2~>~0.5 \cite{Hainline2016ApJ...832..119H}.
Using a limit of W2$-$W3$\leq$3, and W1$-$W2$\leq$0.5 we define a region to select normal galaxy colours. 
This gives us 264 (8\%$\pm$0.5\%) of our radio AGNs whose infrared spectra are dominated by galaxy colours and would not have been identifed as hosts of AGNs using mid-infrared colours only. 
Thus a very high fraction (92\%$\pm$1.7\%) of sources selected in our high frequency compact radio sample are correspondingly identified as AGNs based on mid-infrared colours.

\begin{center}
\begin{figure}
\includegraphics[scale=0.43, angle=0]{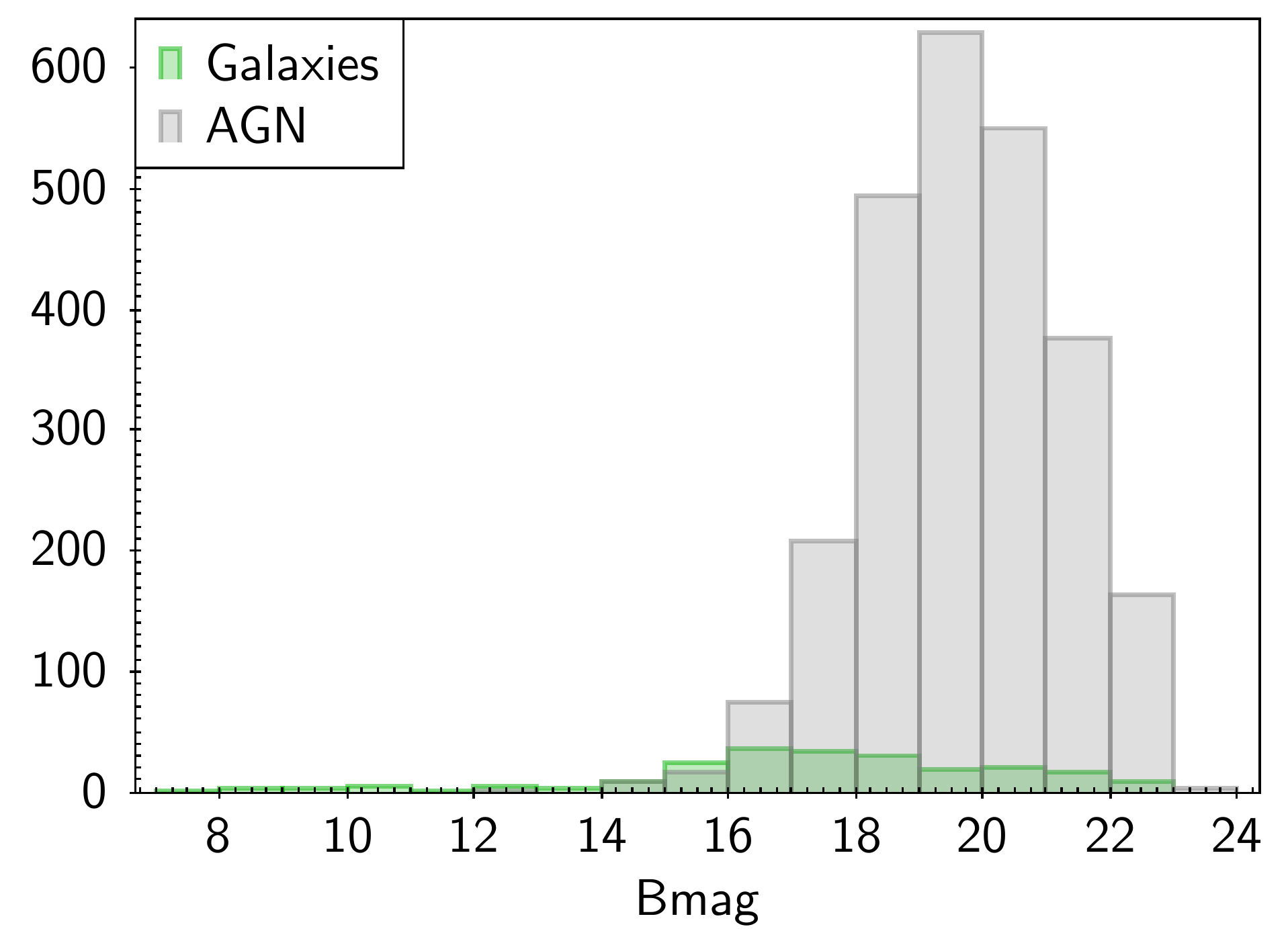}
\caption{Comparison of histograms of B-magnitudes for the AGNs and Galaxies defined in the text.}
\label{Fig:histo-gal-AGNs}
\end{figure}
\end{center}

\begin{center}
\begin{figure}
\includegraphics[scale=0.43, angle=0]{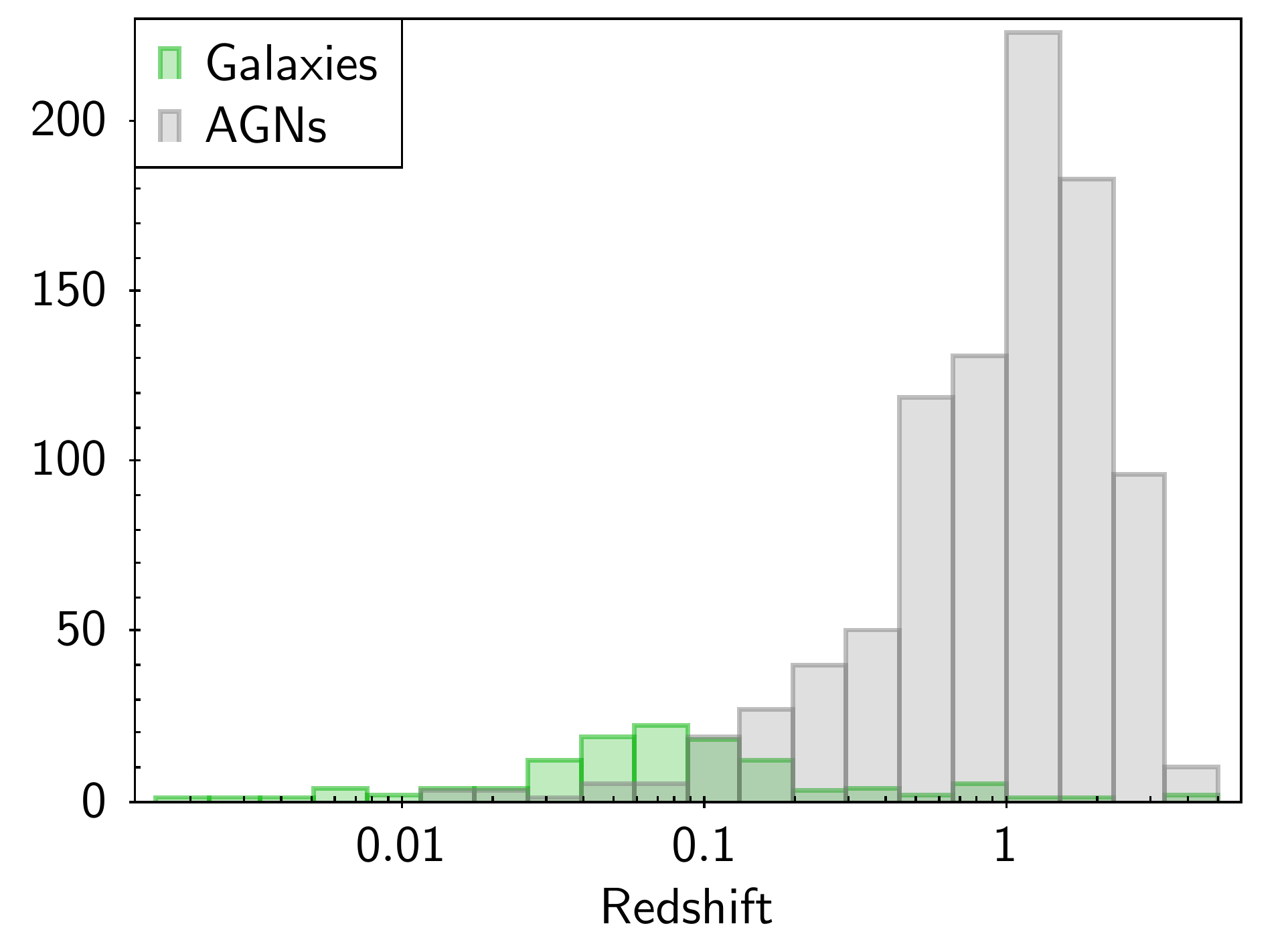}
\caption{Histograms compare the distribution of redshifts for the galaxy and AGN populations.}
\label{Fig:histo-z}
\end{figure}
\end{center}

In \autoref{Fig:histo-gal-AGNs}, we plot the histograms of the B magnitudes of the sources identified as galaxies above using \autoref{Fig:color-color_overlaid}, and compare it against that of the rest of our sources. 
%
The WISE colour-selected galaxies and AGNs have distinct distributions in B magnitude, peaking at 16 and 19, respectively.
In \autoref{Fig:histo-z}, we plot the redshift distributions of these two populations. It shows that with the exception of very small number of high redshift objects, possibly misclassified due to our simple selection criteria, most of the MIR colour-selected galaxies are confined to z<0.2.  
\cite{Ching2017_mnras_464} classified z<0.8 radio sources selected at 1.4 GHz into high and low excitation radio galaxies (HERGs and LERGs respectively, see \cite{Heckman2014_araa_52} for review). They find that at low redshifts the total radio population is clearly dominated by the LERG population (83 per cent) and exhibit infrared colours similar to those seen for these objects. 

Thus, it is most likely that these MIR colour-selected galaxies, which are found to be at low redshifts, are the high radio frequency counterparts of the LERG population. Their compactness indicates that these are the cores of the LERG sources. Since this work focuses on the compact source population only, this is probably only a part of the larger picture. A study of the extended sources to study their distribution in MIR colours will be necessary to provide the overall picture of the compactness of the LERG population at high radio frequencies.
%

\begin{center}
\begin{figure}
\includegraphics[scale=0.55, angle=0]{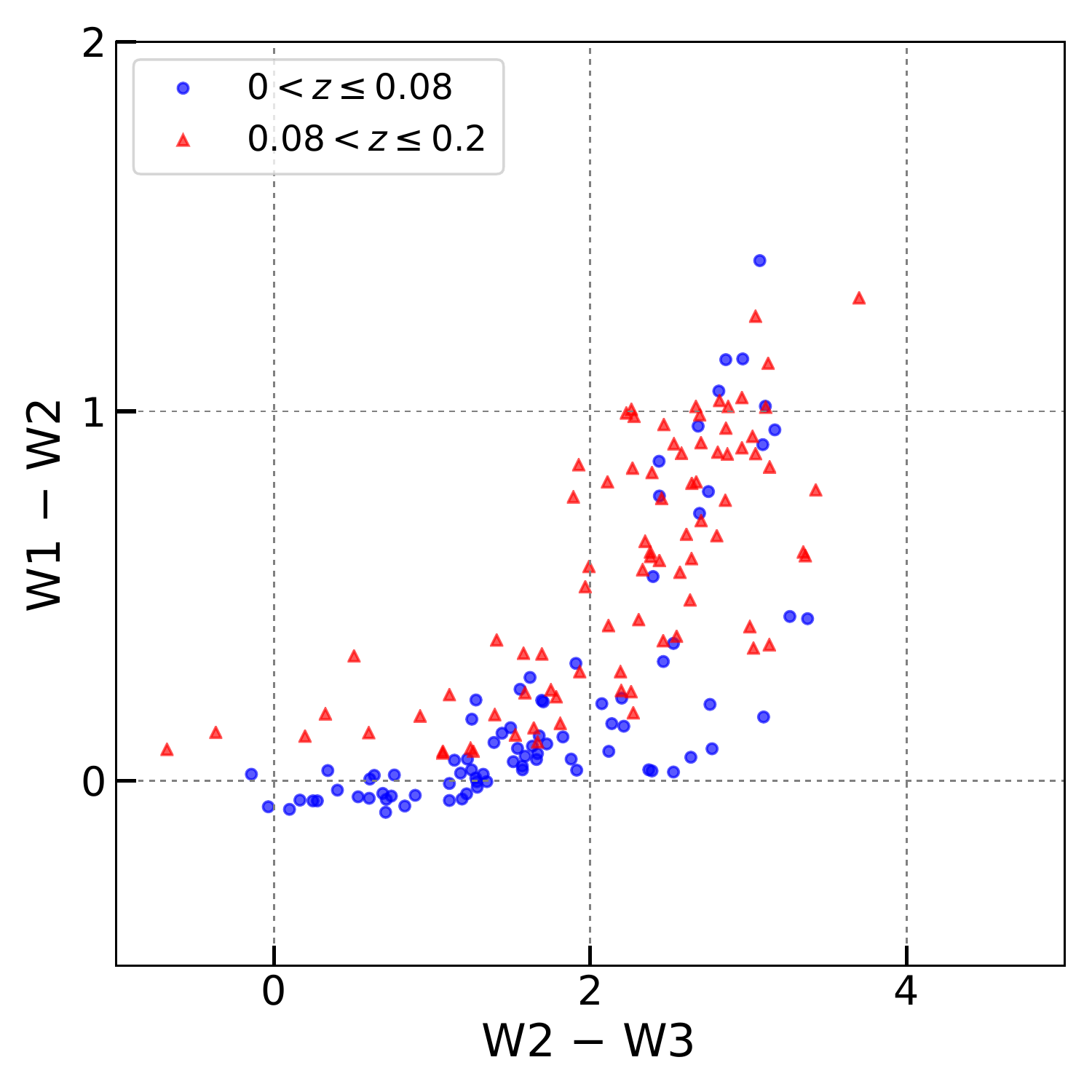}
\caption{Figure shows colour-colour plot for sources with redshift less than 0.2 from our sample, divided into two bins of equal number of sources. Blue filled circles show sources with redshifts between 0 and 0.08, and red filled triangles show sources with redshifts between 0.08 and 0.2.}
\label{Fig:colour-colour_0-z-0p2}
\end{figure}
\end{center}

\autoref{Fig:colour-colour_0-z-0p2} shows the WISE colour-colour plot for all sources below the redshift of 0.2, separated into two bins (0$<$z$\leq$0.08 and 0.08$<$z$\leq$0.2) with similar numbers in both bins (85 and 82). 

\begin{center}
\begin{figure}
\includegraphics[scale=0.51, angle=0]{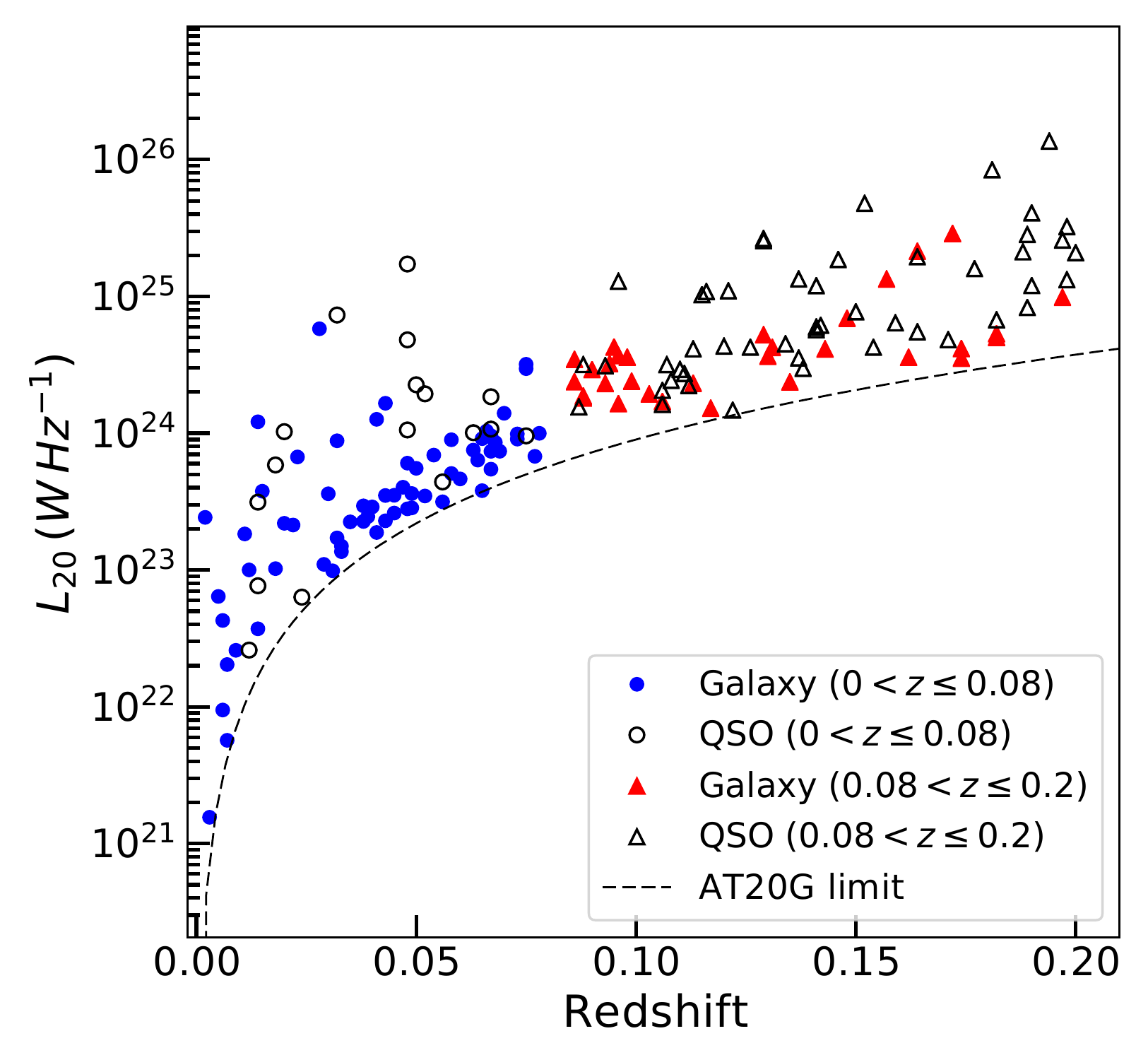}
\caption{Figure shows the luminosity distribution of galaxies (filled circles and triangles) and QSOs (open circles and triangles) as a function of redshift. The distinction between galaxies and QSOs is made using the W1-W2=0.5 limit (see text). Source luminosity is derived using their 20\,GHz flux density. The black dotted line is drawn for luminosities corresponding to the flux density limit of the AT20G survey.}
\label{Fig:z-L20}
\end{figure}
\end{center}

The shallow brightness limits of both AT20G and WISE surveys mean that they select more luminous AGNs with increasing redshifts. Therefore, we expect to see the MIR colours of sources to be increasingly dominated by QSO colours with redshifts. This is demonstrated by the sources in the 0.08$<$z$\leq$0.2 bin showing larger values in the Y-axis compared to those of sources in the 0$<$z$\leq$0.08 bin. Interestingly, even in the lower redshift bin we find that a small number of sources are entirely dominated by QSO colours in mid-infrared wavelengths. These are strong AGNs whose mid-infrared colours dominate over their host galaxy colour contributions. In \autoref{Fig:z-L20}, we plot their 20\,GHz luminosities against redshifts. It shows that for sources in our low-redshift bin, the 20\,GHz luminosity is a good predictor that more luminous radio sources are identified with QSO colours (using our W1$-$W2>0.5 limit) in mid-infrared wavelengths. We also find that when we use a L$_{20}>$5$\times$10$^{24}$ W/Hz threshold, 78\%$\pm$14\%  of compact radio AGNs with redshift below 0.2 are identified with QSO colours in the mid-infrared. Radio luminosity has been widely used to classify AGN types \citep[e.g. see][and references therein]{Heckman2014_araa_52}, however, we note that our estimation of the luminosity uses the flux arising from the compact cores only while most past studies at lower radio frequencies have estimated luminosity using combined flux arising from the compact core and the extended components (i.e. jets/lobes).

\cite{Sadler2014} have used W2-W3\,=\,2 as the demarcation between WISE ``early-type'' and ``late-type'' galaxies (for WISE elliptical and WISE spiral galaxies respectively). When we use this criterion, we find that the compact AGNs with galaxy-like WISE colours are almost equally distributed, with 51.1\%$\pm$4.4\% having the colour characteristics of early-type galaxies and 48.7\%$\pm$4.3\% having the colour characteristics of late-type galaxies. 

\subsection{WISE colours of compact steep-spectrum sources}
We now focus our analysis on an interesting subset of the compact radio source population: the compact steep-spectrum (CSS) sources. In the literature, CSS sources are hypothesized to be early stages of radio galaxy evolution making up approximately 30 per cent of the radio source population with linear sizes $\lesssim$\,15\,kpc \citep{O'Dea1998}, and more evolved than the gigahertz peaked-spectrum (GPS) sources that are confined to even smaller linear scales ($\lesssim$\,1\,kpc). 
In the sample of 33 CSS sources in \cite{O'Dea1998}, they find that 10 sources have linear sizes smaller than 1\,kpc. 
CSS sources are also of interest since these include the ultra steep-spectrum sources, which are viable candidates to identify sources at very high redshifts \cite[e.g.][]{Saxena_2018arXiv180601191S}. The visibility-spectra diagnostic plot in \autoref{Fig:visib-spectra} very conveniently identifies these rare objects. 8.3 per cent of the total AT20G sources can be classified as CSS sources($\alpha_{1}^{20}<-0.5$), $\sim$one-third the fraction stated in \cite{O'Dea1998}. This corresponds well with the linear size distribution in \cite{O'Dea1998}, and is almost certainly because the angular size of 0.15 arcsec corresponds to $\sim$1\,kpc for a wide range of redshifts \citep[$\sim$0.7\,--\,5 figure 1,][]{Chhetri2013}, thus identifying extremely compact steep-spectrum sources. 
Due to the dependence of linear size on redshift, below the redshift of $\sim$0.7 the sources in the AT20G sources will have larger than 1\,kpc linear size.

Using the cut-off of $\alpha_1^{20}= -$0.5, we find 387 (11\%$\pm$0.5\%) sources in our sample of 3610 compact sources that can be considered CSS sources.  
341 (88 per cent) of these 387 steep-spectrum sources have a match in the WISE catalogue, which we can use to investigate their infrared luminosities and colours.

  \begin{center}
  \begin{figure}
  \includegraphics[scale=0.42, angle=0]{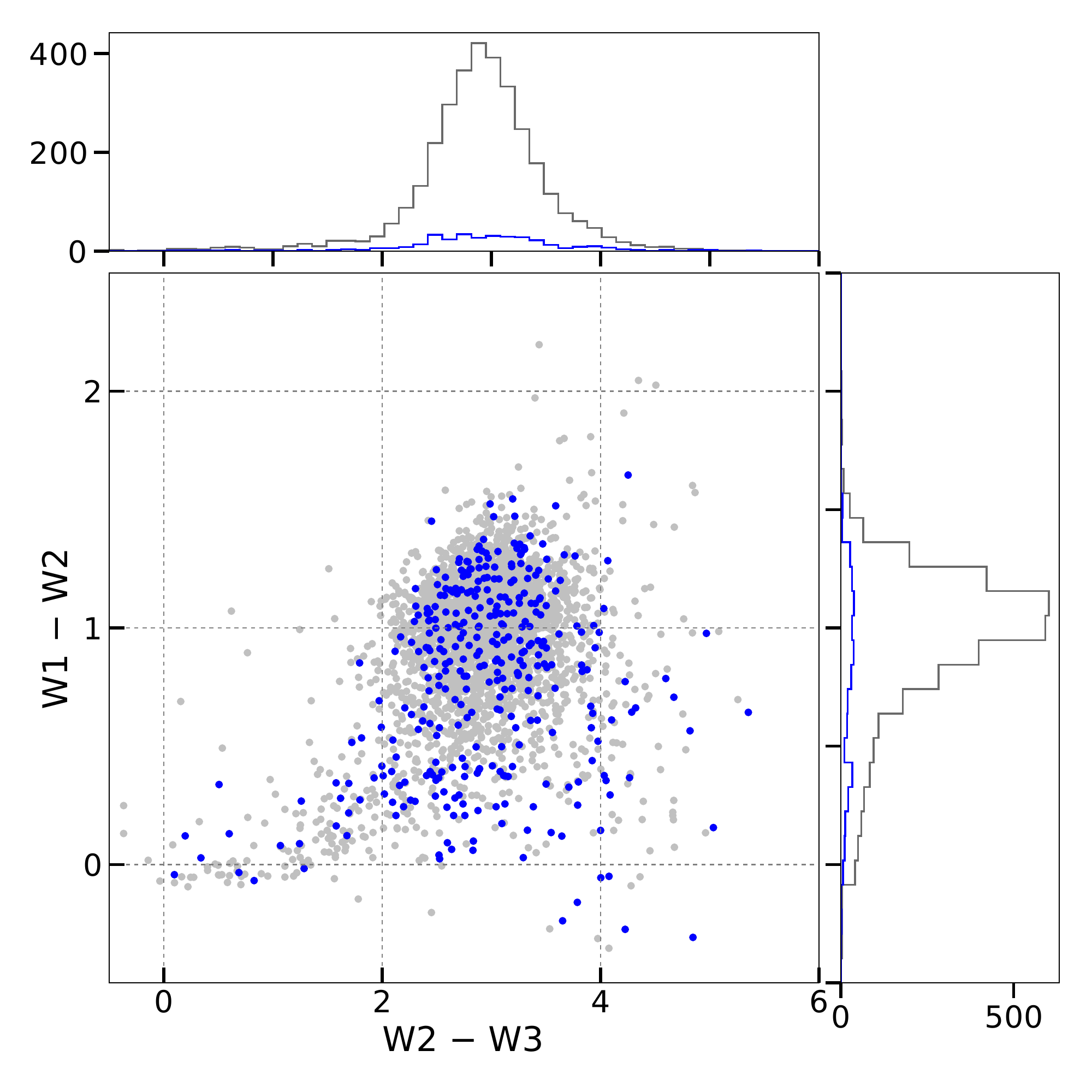}
  \caption{Plot shows the distribution of CSS sources in the colour-colour diagram (blue filled circles). The colours of all compact sources are shown in grey filled circles for comparison.}
  \label{Fig:color-color_CSS}
  \end{figure}
  \end{center}

\autoref{Fig:color-color_CSS} shows a colour-colour plot of CSS sources overlaid on that of the underlying compact source population. Of the 341 sources, 279 (82\%$\pm$5\%) exhibit colours that are not dominated by their hosts galaxies. \autoref{Fig:color-color_CSS} also shows that when the source colours in mid-infrared are not dominated by their host galaxy colours, these objects are indistinguishable from the underlying compact source population which is dominated by the flat-spectrum population. We have redshift information for 85 (approximately one-third) of these objects, with median redshift of 0.82.

An alternative to the hypothesis of CSS sources being young stages of radio galaxies is the `frustrated' scenario where these sources are small because they are confined to smaller spatial scales due to high nuclear density in their hosts \citep{vanBreugel1984_aj_89, Bicknell1997_apj_485}. There are 62 CSS sources whose colours are dominated by the colours of their host galaxies, and the sub-sample for which we have redshift information (27 sources) have median redshift=0.13. \cite{Ching2017_mnras_464} present evidence that the low radio frequency selected sample (as opposed to our high frequency selected compact source sample) at low redshifts (z<0.8) is dominated by the LERG population, whose mid-infrared colours are distributed between early and late-type galaxies. They have found a tendency for LERGs to exhibit mid-infrared colours of late type galaxies, when at higher redshifts in their sample, due to contribution from warm dust owing to e.g. low level star-formation activity. When we separate the host colours of our CSS sources into early and late type using the W2-W3 = 2 limit, we find that a larger fraction of CSS sources (67.7\%$\pm$10.5\%) exhibit late type galaxy colours. These late type galaxies have higher star formation activity than the early type galaxies which indicates the presence of higher density of material. Thus, there is some suggestion that these CSS sources are preferentially found in hosts with higher density of material, and may provide evidence for the frustrated scenario for the small scales of CSS sources.

\subsection{Redshift dependence of AGN SED}
\label{Sec:SED_z_dependence}
Most AGN spectral energy distributions (SEDs) in the literature are empirically derived as the physical processes governing AGN emissions are not quantitatively understood \cite[e.g.][]{Elvis1994, Richards2006, Shang2011}. These empirically obtained SEDs can be complicated by contamination from the host galaxy SED \citep{Assef2010}. 
Due to the relatively high brightness limits of both the AT20G and WISE surveys our work selects luminous AGNs, to very high redshifts. Therefore, our work can identify objects for which the MIR emission is dominated by AGN activity.
With information on only four bands in the infrared, it is not possible to produce detailed SEDs for individual sources. However, using our sample with over a thousand objects for which we have redshift information we are in a position to investigate statistical mid-infrared SED, using colours of radio-selected AGNs, and compare against existing SED templates. 

\begin{center}
\begin{figure}

\includegraphics[scale=0.41, angle=0]{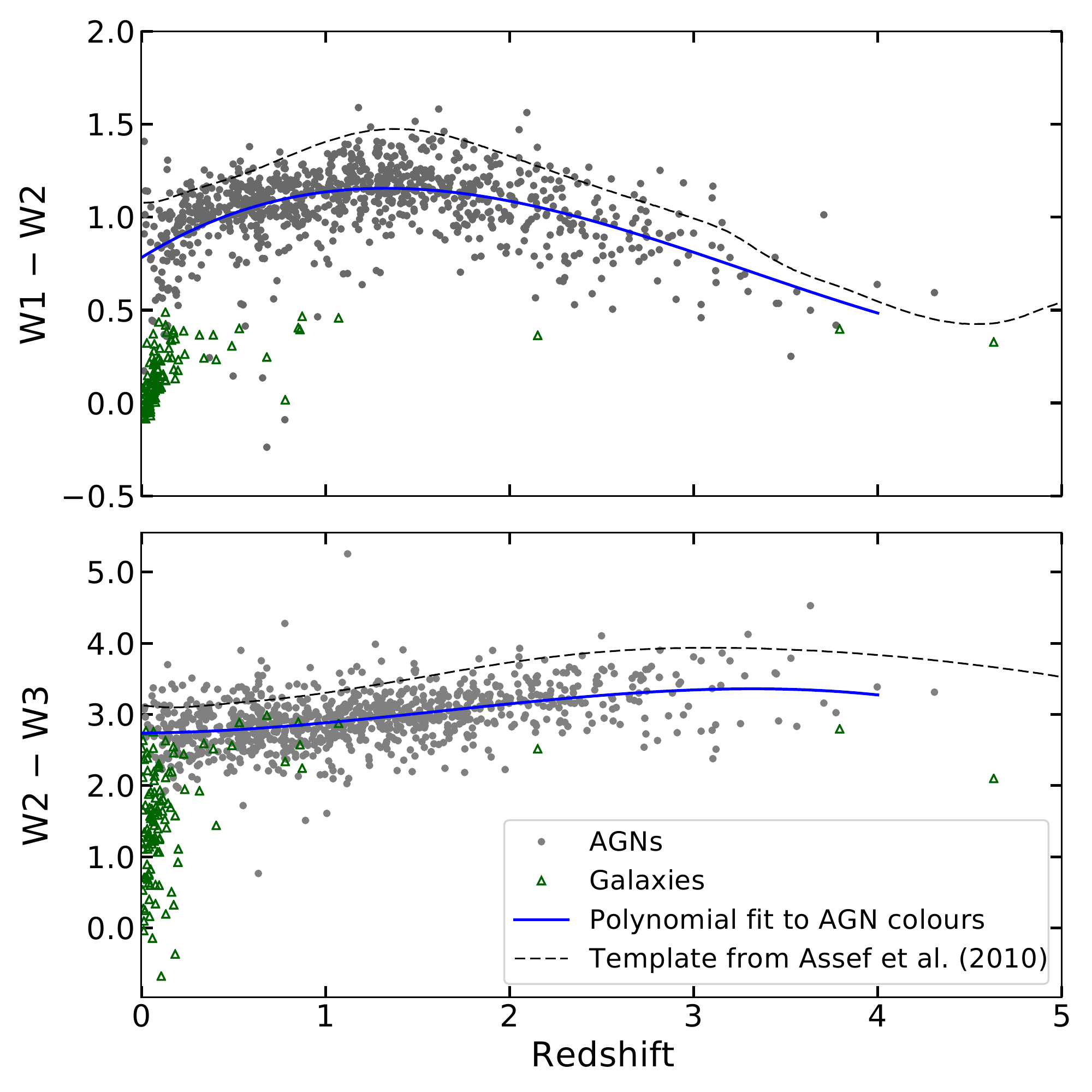}
\caption{WISE colours (the top panel W1-W2 and the bottom panel W2-W3) as a function of redshift for AT20G QSOs (grey filled circles) and galaxies (green open triangles). Third-degree polynomial fits to the distribution of AT20G QSOs are shown with solid blue lines. WISE colours based on template SED of AGNs presented in \protect\cite{Assef2010} are shown with grey dotted line. }
\label{Fig:compareAssef}
\end{figure}
\end{center}

In the top panel of \autoref{Fig:compareAssef}, we present the distribution of mid-infrared colours between 3.4 and 4.6 micron bands as a function of redshift for AT20G sources. The radio AGNs identified to be dominated by their host galaxy colours appear to be mostly confined at redshifts below 0.4. A small number of these sources are also seen at very high redshifts. It is most likely that these objects have been mis-classified as galaxies due to our simple selection criteria for galaxies, and that they truly belong to the QSO category. 

For the sources identified as QSOs, a distinct maximum is seen in the colours at z\,$\sim$\,1.5 suggesting that the sources are $\sim$\,1.3 times brighter in the longer wavelength band  
compared to at redshift $\sim$\,0. Sources then progressively exhibit lower values for W1-W2 colours with increasing redshifts.

We fit a three-degree (cubic) polynomial to W1-W2 colours of AGNs, as a function of redshift. We have limited this fit to within the bounds of within the bounds of $0.0<\,z\,\leq\,4.0$ as our sample becomes very sparse beyond redshift of 4. The resulting fit is shown with a blue solid line in the top panel of \autoref{Fig:compareAssef} and has the form \[ W1-W2 = a_{0} + \sum_{i=1}^{3} a_{i} z^{i}  \] where the values for the constants are $a_{0}$=0.783$\pm$0.022, $a_{1}$=0.612$\pm$0.055, $a_{2}$=-0.289$\pm$0.039 and $a_{3}$=0.029$\pm$0.008. We also plot colours derived using a template SED for AGNs expected in the WISE catalogue by \cite{Assef2010}\footnote{For direct comparison, we converted their AB magnitudes to Vega magnitudes using offsets provided at \url{http://wise2.ipac.caltech.edu/docs/release/allsky/expsup/sec4_4h.html}.}. Their template is derived using over 5000 AGNs selected via different criteria in infrared, optical, X-ray and radio, but with the lowest priority given to radio. To first order our fit has a similar shape to their template; however, the offset in colour is clear. Even with the best estimate of offset in magnitudes of 0.20 to match the two profiles, we find that the differences between the two profiles are significant. A similar fit to W2-W3 colours is presented in the bottom panel of \autoref{Fig:compareAssef}, and we once again see a similar offset in magnitudes between our fit and that provided by \cite{Assef2010} template. The parameters of this fit are $a_{0}$=2.740$\pm$0.042, $a_{1}$=0.018$\pm$0.107, $a_{2}$=0.158$\pm$0.075 and $a_{3}$=-0.032$\pm$0.015. 

In \autoref{Fig:color-color_overlaid}, we notice an offset in the centroid of the radio AGN population towards lower values in both X and Y axes. From the \citeauthor{Wright2010} plot, the expected central value of the bubbles representing the Seyfert galaxies and QSOs is at colour coordinates of 3.5, 1.25. We find that the centroid of our AGN population lies at 3.0, 1.0. 

The \cite{Assef2010} template shown in \autoref{Fig:compareAssef} was made using 10 times as many objects, selected using non-radio (IR, optical, X-ray) selected AGNs compared to radio selected sources. Therefore, it is natural that their template has a stronger influence from the SED of the larger infrared/optical selected population. 
%
Our technique selects the most powerful (radio) AGNs, which are hosted in massive galaxies \citep[e.g.][]{Matthews1964, Urry1995}. It is then possible that the difference we see in \autoref{Fig:compareAssef} can be explained by this selection of powerful radio AGNs in which the stronger contributions of the hosts to the total MIR emission causes the colours of the radio AGNs to be less red in comparison to the \cite{Assef2010} sources, which are mostly radio quiet. Even though we have not attempted to disentangle this effect of host galaxy contributions to the total source SED, this work contributes towards a template SED for the very powerful and compact radio selected AGNs.

\begin{figure*}
        {\label{SubFig:color_z_dance1}
        \includegraphics[width=.4\linewidth]{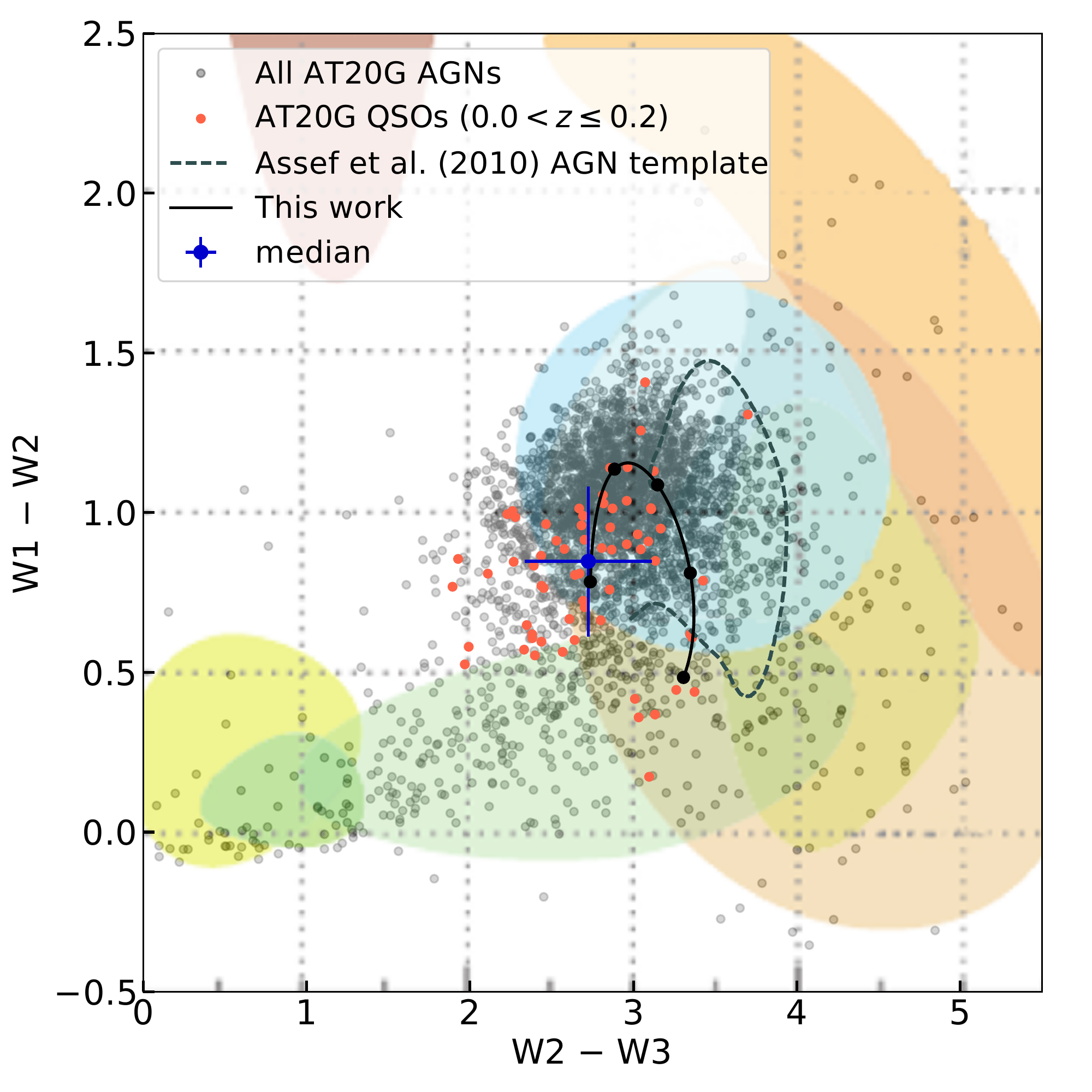}} \quad
        {\label{SubFig:color_z_dance2}%
        \includegraphics[width=.4\linewidth]{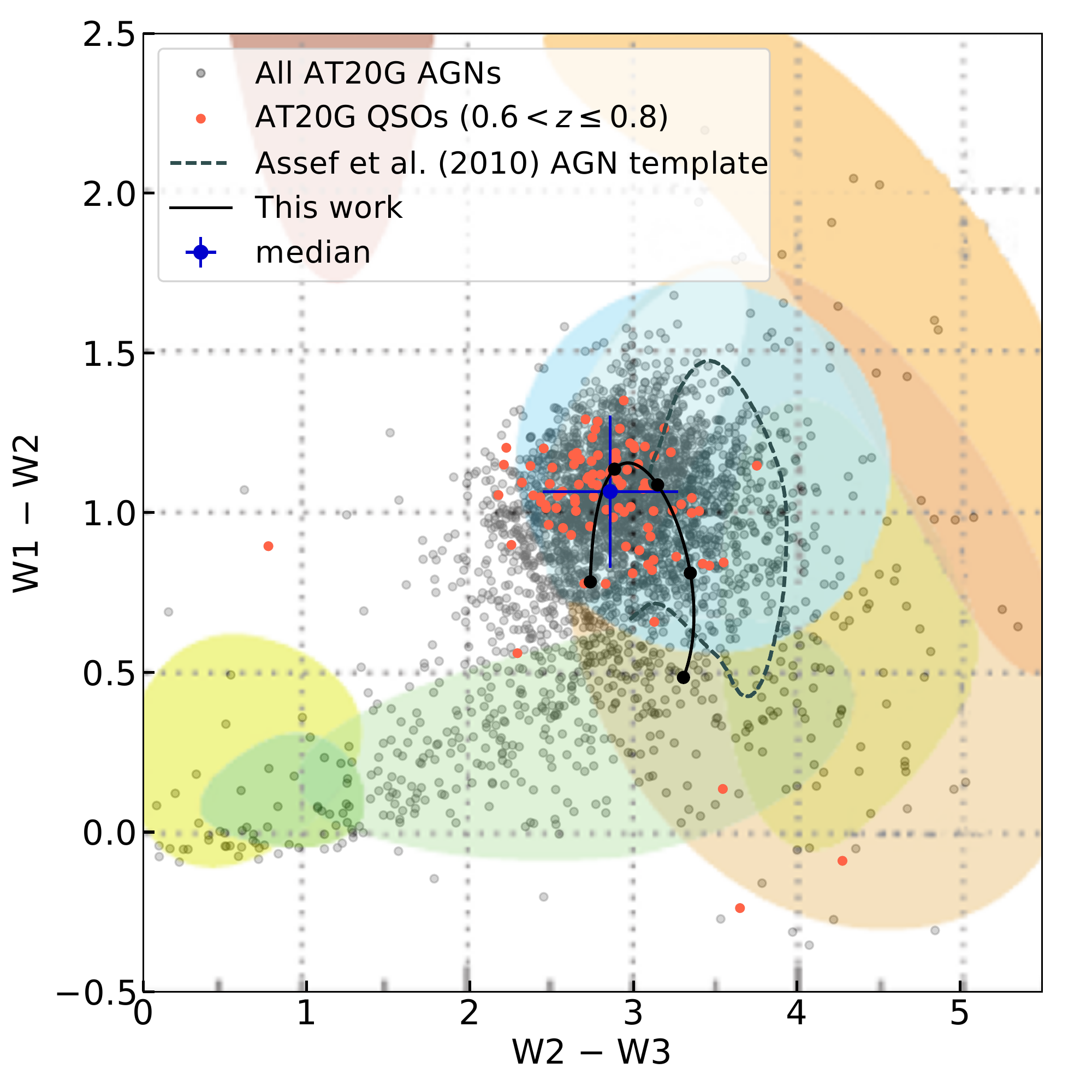}}
\\
        {\label{SubFig:color_z_dance3} 
	\includegraphics[width=.4\linewidth]{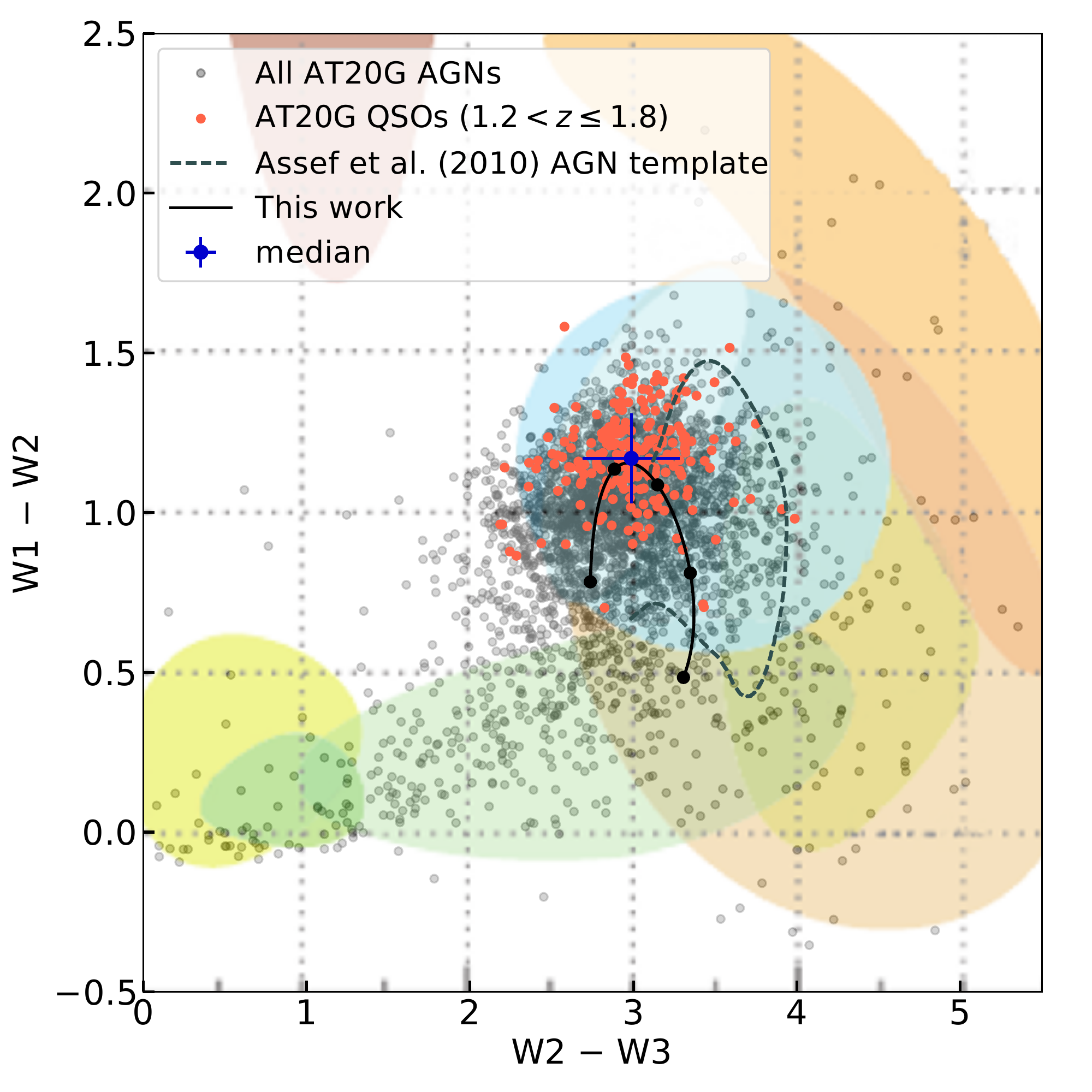}}
        {\label{SubFig:color_z_dance4}        
        \includegraphics[width=.4\linewidth]{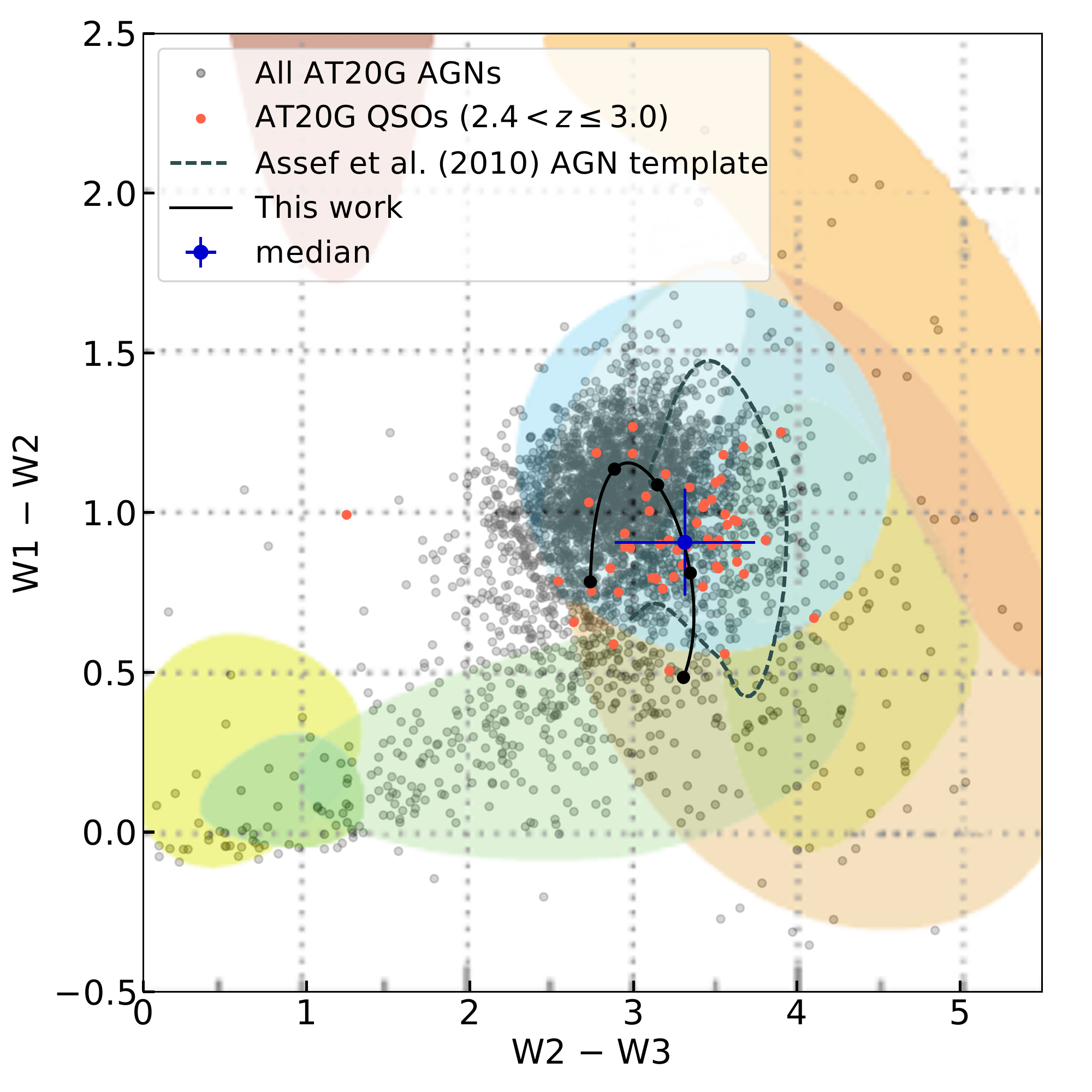}}

        \caption[]{The distribution of the AT20G QSOs as a function of redshift on the WISE colour-colour plot. The four representative panels show sources in different redshifts bins (orange filled circles) in increasing order or redshift. Blue filled circles, with one standard deviation errorbars, show the median colour for all AT20G QSOs in the corresponding redshift bin.  The path followed by our fit (see text) is shown with a black line, with black filled circles marking redshifts 0, 1, 2, 3 and 4 respectively. The AGN template from \protect\cite{Assef2010} is shown with a black dotted line. Dark circles in the background show all AT20G sources for reference. An animated movie to show the movement of sources as a function of redshift in the colour-colour plot, with contiguous redshift bins, is available as supplementary material with this paper.}

\label{Fig:color_z_dance}
\end{figure*}

Using the polynomial fits discussed above, we plot a locus followed by radio selected AGNs in the WISE colour-colour plots as a function of redshift. This is presented in \autoref{Fig:color_z_dance}. The path taken by radio selected AGNs is much more confined compared to that defined by the generic AGN template by \cite{Assef2010}, possibly due to contributions of the host galaxy as discussed above. In the sub-figures of \autoref{Fig:color_z_dance}, we show the displacement in WISE colours for radio selected AGNs as a function of redshift. Thus, with a SED based on compact radio selected AGNs, it may be possible to improve estimates of photometric redshifts for powerful radio AGNs \citep[e.g.][]{Salvato2009}.

\subsubsection{Non-detections in WISE as candidate high redshift sources} 
\label{Sec:non_detections}
Finally, we note that selecting infrared faint sources with small angular size is an excellent method to identify radio sources at very high redshift \citep[z\,>\,4,][]{Ker2012MNRAS.420.2644K}. The relatively shallow observations in WISE means that sources at very high redshifts are undetected. Indeed, approximately a third of our sources with non-detections in WISE have optical identifications and most (91$\pm$11\%) of these sources with optical identifications have B magnitude values\,$\geq$\,19. This shows that the small number of non-detections in our sample are excellent high redshift candidates. These differ from the candidates identified from low radio frequency surveys in that most low radio frequency selected candidates are obtained using an ultra-steep-spectrum criteria. Most of our non-detections exhibit flat radio spectra and are expected to be beamed. If relativistic beaming at radio and mid-infrared are different, i.e., if the radio beam has a wider angle, our radio selection will detect AGNs with amplified radio emission but not as amplified in mid-infrared making them undetectable in WISE. Barring such objects, which can be at not-very-high redshifts, our list of non-detections in mid-infrared is an important complement to the traditionally identified high-redshift radio galaxy candidates. 

\section{Summary}

We selected 3610 compact sources ($<$0.15 arcsec) from the AT20G high-angular-resolution catalogue to make a cross-match against the AllWISE mid-infrared survey catalogue. The selected compact sources are primarily (89 per cent) flat-spectrum cores  of radio galaxies, as expected from that of a high frequency radio survey. 

Using a semi-automated cross-matching technique, we have made a catalogue which has 3300 (91.4\%\,$\pm$\,1.6\%) matches, 91 (2.5\%\,$\pm$\,0.3\%) rejects due to various reasons and 219 (6.1\%\,$\pm$\,0.4\%) non detections in the WISE catalogue. We present the resulting catalogue complete with radio data from the AT20G survey, compactness information, optical identiciation with redshift, B and R magnitudes when available, and the WISE magnitudes at 3.4, 4.6, 12 and 22 micron wavelengths. 

3036 of the compact AGNs with WISE cross-matches exhibit QSO/Seyfert colours in mid-infrared wavelengths (with W1-W2$>$0.5), which is a significant majority (92.0\%\,$\pm$\,1.7\%) of our sample. Only a minority ($\sim$8.0\%\,$\pm$\,0.5\%) exhibit colours that are dominated by their host galaxy colours. 
Therefore, our sample of high frequency compact sources has a very high rate of identification with MIR QSOs.

In our sub-sample of low redshift compact sources (z < 0.2), radio luminosity (L$_{20}$) is a good indicator to identify QSOs, with 78\%$\pm$14\% of sources with L$_{20}\,>$\,5$\times$10$^{24}$ W/Hz exhibiting mid-infrared QSO colours. We note that for these sources, the luminosity is due to the  flux arising from the compact cores only as opposed that for low frequencies selected sources, whose luminosity is due to combined flux arising from the compact core and the extended components (i.e. jets/lobes).

We investigated the sample of compact steep-spectrum sources using their mid-infrared colours conveniently identified using the AT20G visibility-spectra diagnostic plot. We find that 8.3 per cent of AT20G sources are CSS sources, which is approximately one-third of the number expected from the literature. This is possibly because our angular size criteria identifies objects <1 kpc across a wide redshift (z$\sim$0.7 to 5) range, selecting extremely compact steep-spectrum sources. We find cross-matches for 88 per cent of CSS sources in our sample. Of these, 82\%$\pm$5\% of CSS sources exhibit colours of QSOs, which are indistinguishable from the compact flat-spectrum source population using mid-infrared colours only. These are found to be at moderate redshifts, with a median redshift of 0.82.
The remaining CSS sources exhibit their mid-infrared colours to be dominated by the colours of their hosts. These are found at lower redshifts with median redshift of 0.13. A higher fraction of CSS sources (67.7\%$\pm$10.5\%) exhibit colours in MIR which is characteristic of late-type galaxies, in comparsion to the fraction of overall high frequency compact sources exhibiting MIR galaxy colours (48.9\%$\pm$4.3\%). This indicates some level of star forming activity in the CSS hosts and, therefore, a higher density of gas in the central ISM. Thus, our data provides some evidence that the small size of the CSS sources is due to their confinement in hosts with higher density regions.

Using polynomial fits made to the WISE colours of our compact radio AGNs, we make comparisons against SEDs of AGNs selected primarily using infrared/optically techniques. From these comparisons we note that the SED for compact, high frequency selected radio AGNs is less red compared to that for AGNs selected primarily using optical techniques or infrared colours. This is possibly due to contributions in the mid-infrared from the (massive) hosts of radio AGNs preventing the AGN colours from becoming redder still. 
Thus, we propose that the use of a mid-infrared SED template purely derived for radio sources may be useful to improve photometric redshift estimates for radio AGNs.

Finally, we note that the small number of AGNs with no counterparts in WISE are excellent candidates for sources at very high redshifts. Since these objects are selected from flat-spectrum AGNs, high redshift sources identified from these candidates will provide an important compliment to those identified using the most-successful ultra steep-spectrum criteria.

\section*{Acknowledgements}
The authors wish to thank the anonymous referee for providing valuable comments which improved the quality of this publication. 
The authors also wish thank K.~Marsh for helpful discussions of the WISE data. 
R.~C.~wishes to thank G.~Drouart for helpful discussions.
We used the TOPCAT software \citep[]{Taylor2005} for the analysis of some of our data.
This publication makes use of data products from the Wide-field Infrared Survey Explorer, which is a joint project of the University of California, Los Angeles, and the Jet Propulsion Laboratory/California Institute of Technology, funded by the National Aeronautics and Space Administration.
The Australia Telescope Compact Array is part of the Australia Telescope National Facility which is funded by the Australian Government for operation as a National Facility managed by CSIRO. This publication makes use of data products from the Wide-field Infrared Survey Explorer, which is a joint project of the University of California, Los Angeles, and the Jet Propulsion Laboratory/California Institute of Technology, funded by the National Aeronautics and Space Administration.


\bibliographystyle{mnras}
\bibliography{references-at20g_wise}

\bigskip
This document has been typeset from a \TeX/\LaTeX file prepared by the author.

\appendix

\section{Deficiency of nearby WISE counterpart numbers}\label{appendix:Dearth}
As an \textit{a priori} step for performing the cross-matching between AT20G and the AllWISE catalog, we performed a random comparison of ``nearest neighbors" as a function of separation; i.e., counts of nearest neighbors to AT20G sources compared to counts of nearest neighbors to random positions on the sky. We created the catalogue of random positions by shifting each AT20G position by one full degree in Galactic longitude. For any roughly homogeneous-density catalogue, the histogram of ``random" neighbors will be a linear function rising from zero at the origin, owing to the linear increase in area; this distribution represents the foreground/background source density. In theory, the distribution of ``real'' neighbors should be a function that is sharply peaked near the origin, drops with distance to a local minimum, then rises linearly for large separations; this distribution represents the sum of true counterparts plus the random foreground/background source density.

To our initial surprise, the distribution of nearest WISE neighbors around AT20G sources shows a deficiency of counterparts in the 3--13 arcsecond range with counts that drop below the expected distribution from random matching (see \autoref{Fig:astrometry}). Similar behavior was noted in \cite{Krawczyk2013}, where the claim was made that this effect is ``due, in part, to the large beam size of WISE (FWHM = 6$^{\prime\prime}$)" but no further explanation was offered.

The explanation for this effect is not that the foreground/background sources (which must be present) are actually missing, but rather that they have become blended with the true WISE counterpart. Bright WISE sources can saturate nearby pixels, or can otherwise be so bright that nearby fainter sources become indistinguishable. This blending effect happens all over the sky whenever a faint source (typically with B mag fainter than 19) happens to lie close to the line-of-sight towards a very bright source. The WISE survey applies a de-blending algorithm to distinguish blended sources, but the minimum resolvable separation can increase in the case of a faint source within the wings of a much brighter one (K. Marsh, private communication). We therefore may expect the ``deficiency of random neighboring sources" effect to become more pronounced for brighter sources.

We verified this behavior using a similar random nearest neighbor comparison for Sloan Digital Sky Survey quasars (SDSS) (Schneider et al. 2010) as shown in \autoref{appendix:fig:comparison_bright-weak_AGNs}. SDSS quasars are identified partly by being optically luminous, and are typically luminous in the mid-infrared as well. Therefore, their WISE counterpart is often blended with one or more fainter foreground/background WISE sources, leading to a "deficiency of random neighboring sources" effect. To test whether the effect is more pronounced for brighter sources, we performed the analysis separately for quasars classified as ``bright" ($<$ 18.5 mag in all SDSS bands) and ``faint" ($>$ 20 mag in all SDSS bands). As shown in \autoref{appendix:fig:comparison_bright-weak_AGNs}, the deficiency of random neighboring sources is much stronger for the bright subset; the deficiency persists beyond 15 arcseconds, demonstrating that WISE counterparts to bright SDSS quasars can have wings extending well beyond 15 arcseconds from the point source position. The effect is much less for the faint quasar subset, but still persists nearly to 10 arcseconds -- ``faint" quasars are still very bright in the mid-infrared!

As a more direct test, we separated the SDSS quasars into subsets based on the properties of their nearest WISE neighbor rather than their optical brightness. \autoref{appendix:fig:dearth-bright_weakQSO} shows the results for WISE counterparts that are ``bright" or ``faint" in all four WISE bands; the numbers are much less because these samples only include WISE counterparts that are detected in all four WISE bands. Again, the results show that the effect is more pronounced for bright WISE sources. Finally, we divided the sample into subsets based on whether the WISE counterpart to the SDSS quasar is a point source (WISE parameter ext\_flg=0), an extended source (ext\_flg=1), or possibly an extended source (ext\_flg=2). The majority of WISE counterparts to SDSS quasars are point sources in WISE, so the top panel of \autoref{appendix:fig:dearth-ext_compactQSO} is similar to the top panel of \autoref{appendix:fig:comparison_bright-weak_AGNs}. The effect is much less significant for the extended WISE sources, perhaps partly because the subset is much smaller, but also perhaps because the extended sources have a lower surface brightness and are easier to distinguish from foreground/background point sources.

\begin{center}
\begin{figure}
\includegraphics[scale=0.57, angle=0]{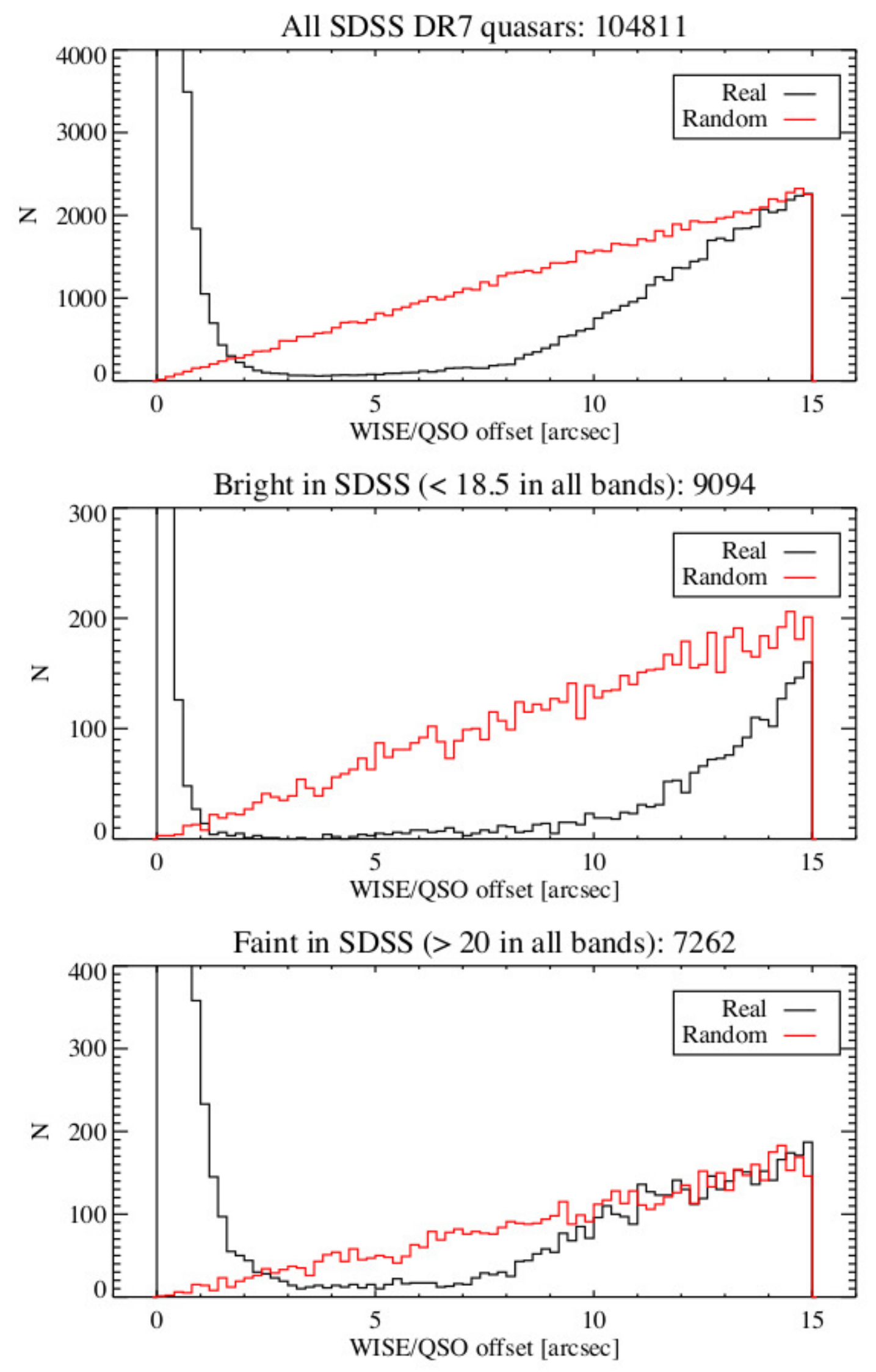}
\caption{Plot compares the effects of deblending between bright and weak QSO, selected from SDSS, in finding the cross-match for sources.}
\label{appendix:fig:comparison_bright-weak_AGNs}
\end{figure}
\end{center}

\begin{center}
\begin{figure}
\includegraphics[scale=0.57, angle=0]{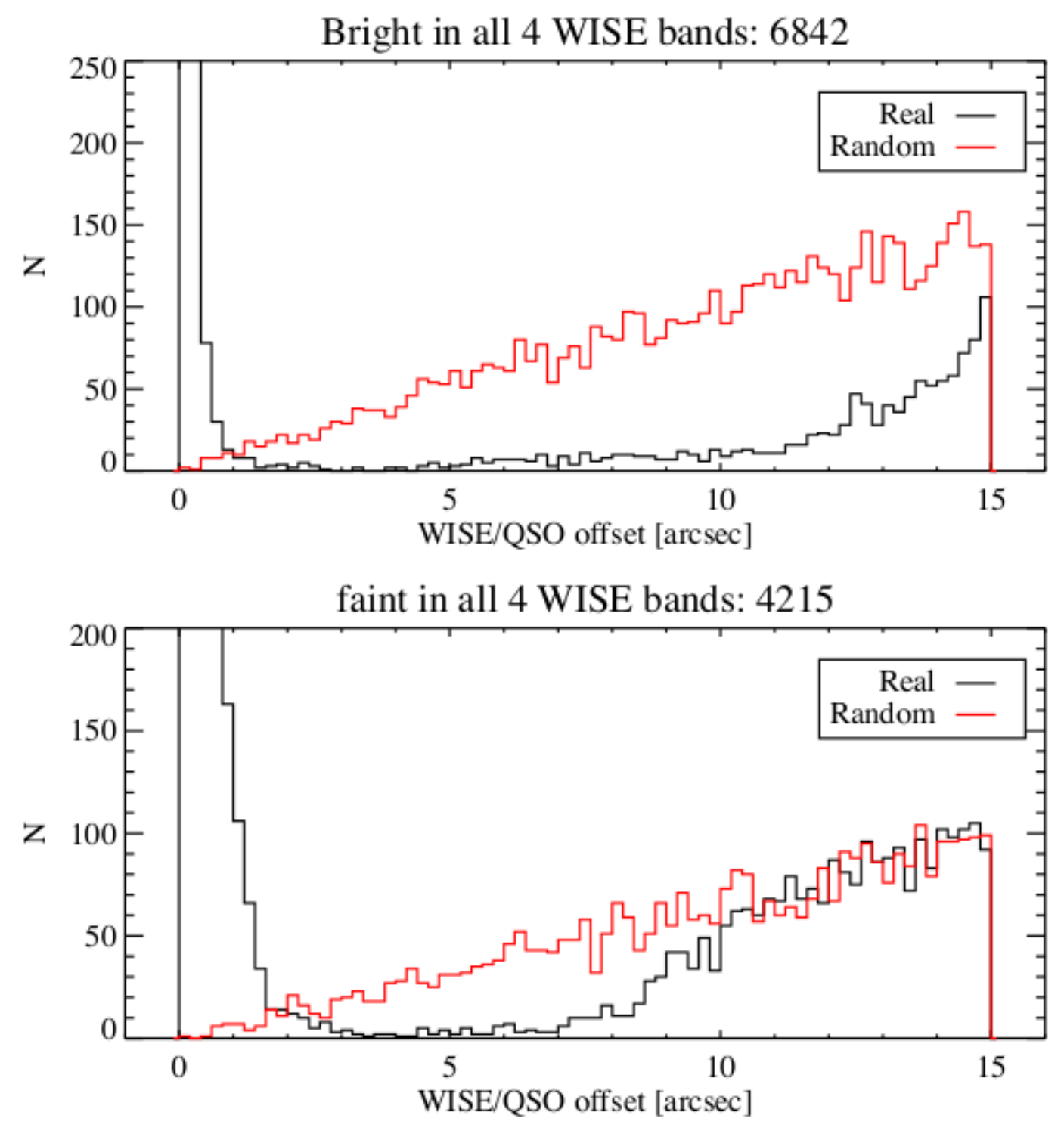}
\caption{Plot compares the effects of deblending between bright and faint objects in all WISE bands.}
\label{appendix:fig:dearth-bright_weakQSO}
\end{figure}
\end{center}

\begin{center}
\begin{figure}
\includegraphics[scale=0.57, angle=0]{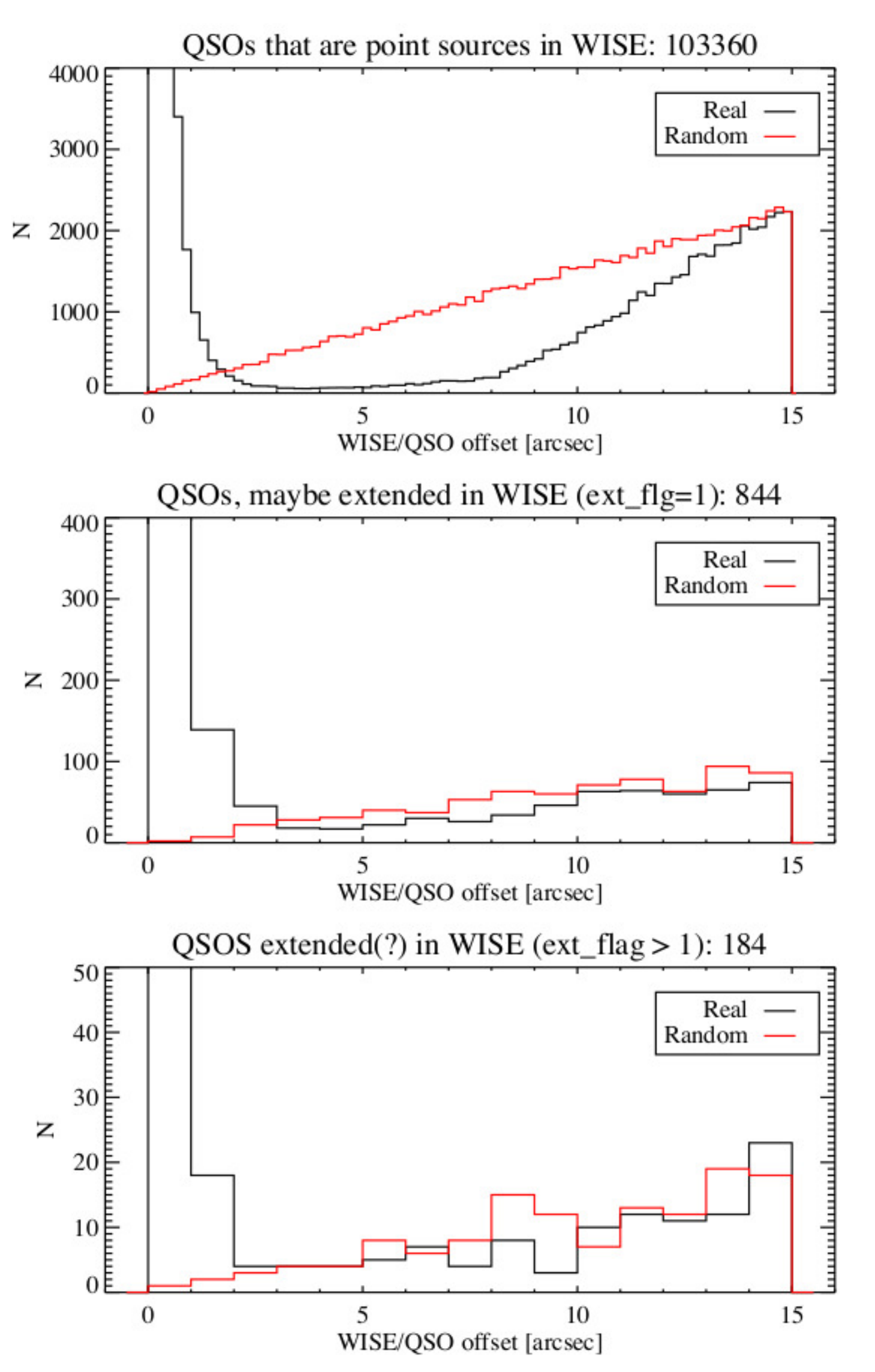}
\caption{Plot compares the effects of deblending between point-like, extended sources and possibly extended source in WISE. }
\label{appendix:fig:dearth-ext_compactQSO}
\end{figure}
\end{center}

\section{Detailed description of flags}
\label{Appendix:Flags}

\subsubsection*{Magnitude type flag}
The measurement and presentation (of flux and, therefore,) magnitude from WISE images is dependent upon the profile of the source. In short, measurements from different sized apertures are used to produce a magnitude profile that is compared against that expected from an unresolved source. In cases of unresolved sources, the use of instrumental profile-fit magnitude is appropriate. For resolved sources, elliptical apertures are used to obtain photometric magnitudes and magnitudes derived from elliptical aperture (``gmag'')  is more appropriate (and is recommended, private communications, T. Jarrett). In between these two extremes, apertures with different radii are used. 
(For our sources, we consider magnitudes from the circular WISE apertures 2-5; apertures 1 to 5 correspond
to 5.5'', 8.25'', 11.0'', 13.75'', and 16.5'', respectively. Aperture 2 with an 8.25 arcsec radius is considered to be the standard aperture for W1.)
In the catalogue we have provided information on which magnitudes are presented as ``Magnitude type'' flag. These are provided as 4 alphanumeric flags, each letter or number representing one WISE band going from left to right for WISE bands 1 to 4.  For sources whose magnitudes are obtained from \cite{Sadler2014}, the letter ``s'' is used.

\subsubsection*{Contaminated and confusion (cc\_flag) flag}
The four character string of the cc\_flag, as given in the WISE supplementary documents\footnote{http://wise2.ipac.caltech.edu/docs/release/allwise/expsup/}, indicates that photometry and/or position measurements of a source may be contaminated or biased due to proximity to an image artifact. Characters from left to right correspond to WISE bands 1 to 4 respectively. \\
``D, d'' - ``D'' may be a spurious detection of and ``d'' may be contamination by a diffraction spike from a nearby bright star on the same image. \\
``P, p'' - ``P'' may be a spurious detection of and ``p'' may be a contamination by a short-term latent image left by a bright source.\\
``H, h'' - ``H'' may be a spurious detection of and ``h'' may be a contamination by the scattered light halo surrounding a nearby bright source. \\
``O, o'' - ``O'' may be a spurious detection of and ``o'' may be a contamination by an optical ghost image caused by a nearby bright source.\\
``0'' - the number zero flag corresponds to sources unaffected by any known artifacts in the observing band. \\
When a source is affected by more than one type of artifact or condition, the cc\_flag value priority in each band is in the order ``D, P, H, O, d, p, h, o, 0''.

\subsubsection*{Photometric quality (ph\_qual) flag} 
The WISE ph\_flag is a four character flag that provides a summary of the quality of the profile-fit photometry measurement as derived from the measurement signal-to noise ratio. Characters from left to right correspond to WISE bands 1 to 4 respectively. We provide descriptions below as provided in WISE supplementary documents.\\
The ``A'' flag is given to sources detected (in this band) with a flux signal-to-noise ratio w$?$SNR$\geq$10, where W$?$ is WISE band W1, W2, W3 or W4.  \\
The ``B'' flag is given to sources detected with a flux signal-to-noise ratio 3$<$w$?$SNR$<$10.\\
The ``C'' flag is given to sources detected with a flux signal-to-noise ratio 2$<$w$?$SNR$<$3.\\
The ``U'' flag is given to sources detected with a flux signal-to-noise ratio w$?$SNR$<$2, and hence the magnitude value is an upper limit on magnitude. The profile-fit magnitude w$?$mpro is a 95\% confidence upper limit.\\
The ``X'' flag is given to sources whose profile-fit measurement was not possible at this location in this band. \\
The ``Z'' flag is given to sources whose profile-fit flux measurment was made at this location but flux uncertainty could not be measured. 

The ph\_qual flag does not apply for sources whose `gmag' has been presented. The histogram of photometric quality for our cross-matched sources is presented in \autoref{Tab:histo_PhQual}. 

\begin{table}
\begin{center}
\begin{tabular}{lllll}
\hline
\textbf{Quality Flag}	& 	\textbf{W1}	&\textbf{W2}	& 	\textbf{W3}	& \textbf{W4} 	\\
\hline 
A	&	3099	&	2887	&	1340	&	440\\
B	&	106	&	308	&	1227	&	1109\\
C	&	2	&	9	&	271	&	451\\
U	&	0	&	4	&	368	&	1206\\
X	&	1	&	0	&	2	&	2\\
Z	&	0	&	0	&	0	&	0\\
\hline
\hline
\end{tabular}
\end{center} 
\caption{Table shows the WISE photometric quality histogram for the cross-matched sources. Descriptions for different flags are given in Section \ref{Sec:presentCatalogue}. For convenience, A = source with W?SNR$\geq$10 where ``?'' is band 1,2,3 and 4, B = 3$<$W?SNR$<$10, C = 2$<$W?SNR$<$3, U = W?SNR$<$2 so magnitude is an upper limit, X = profile fit measurement not possible and Z = profile fit measurement made but uncertainty could not be calculated.}
\label{Tab:histo_PhQual}
\end{table}

\subsubsection*{Multiband quality flag}  
The nature of the WISE data necessitates a complicated system to indicate the quality of magnitudes in the WISE catalogue. In this catalogue of compact radio sources, we present a simplified quality flag to assist in interpreting the data quality. The four numbers in the flag represent the quality in each WISE band from left to right representing bands 1 to 4 respectively. \\
A flag value ``0'' means excellent detection in this band, with photometric quality flag (see below) of either A or B and cc\_flag (see below) of zero. \\
A flag value of ``1'' indicates sources with upper limit photometry on this band (ph\_qual flag of ``U'', regardless of cc\_flag value). \\
A flag value of ``2'' indicates that the magnitude in this band should be used with caution (ph\_qual value is either ``A'' or ``B'' but cc\_flag values is not zero). \\
A flag value of ``3'' indicates bad photometry in the band and should not be used (ph\_qual is either ``X'' or ``Z'' regardless of cc\_flag value). Only one source (J155355-235841) in our catalogue has the flag value of X in W1 only and none have the value of Z.

\subsubsection*{Extended flag (ext\_flg)} 
\label{Appendix:ExtFlag}
The WISE catalogue provides an ``extended flag'' to signify that the source is extended in WISE bands. See the description for each category in the description for columns in \autoref{Tab:main_table}. We find that 88.6\% of AT20G sources with WISE counterparts have an extended flag of 0 and are considered unresolved in WISE. 8.2\% of sources have extended flags of 1, 2 or 3, and are suspected to be resolved or the source is found to be located within the isophotal extent of 2MASS extended source catalogue. 3.2\% of sources of AT20G-WISE cross-matched sources that are compact in AT20G are expected to be extended in WISE. 


\section{Sources with extended structures in AT20G survey}
\label{Appendix:HighlyExtendeSources}
The AT20G catalogue identified 337 sources that were resolved in the $\sim$200\,m baselines of the ATCA during follow-up observations. This corresponds to sources with sizes larger than $\sim$5 arseconds. 25 of these sources are part of our compact source sample (identifiable with a flag `e' in column 10 of \autoref{Tab:main_table}). These sources have extended structures at large scales in addition to the compact AGN core. The overall radio properties of these sources in the AT20G catalogue are a complicated combination of the compact component and the extended structure(s), and we have not attempted to separate the flux arising from the compact and the extended components using the 6-km visibility for this work. For two of these objects, the WISE counterparts were rejected as being blended, and 4 more are associated with 2MASS extended sources. Thus, the observed mid-infrared counterparts for these extended sources may also be due to be a combination of the central AGN and extended host structures. 

\section{Centaurus A}
\label{Appendix:CenA}
The very large scale extended components of the low redshift giant radio galaxy Centaurus A are resolved out in the long baselines of the ATCA at 20 GHz and they are sensitive to the compact core in the nucleus only. Hence, in the AT20G high-angular-resolution catalogue it has a 6-km visibility of 1.00, thus, it is selected as a compact source in our selection criteria. 

The photometric magnitudes for Centaurus A (NGC 5128) were provided by T.~Jarrett. Even though the AT20G high-angular-resolution catalogue is sensitive only to the central AGN of Centaurus A, we believe the WISE characteristics of the galaxy are appropriate for this catalogue and for further analyses.

\section{Thermal sources and incorrectly identified source}
A small number of sources occupy the region designated ultra luminous infrared galaxies (ULIRGs)/LINERS and obscured AGNs. \cite{Chhetri2015} have found that high Galactic latitude and extragalactic thermal sources in the local universe are also effectively identified on the upper right-hand side of this plot. They used the extended nature of these sources in \autoref{Fig:visib-spectra} and their mid-infrared colours to effectively identify thermal sources in the local universe. All known thermal sources identified in AT20G were removed from this analysis. 

The technique presented in \cite{Chhetri2015} may be turned around to discriminate against incorrectly identified objects, as shown in the following section with the case of AT20G J012930-733311. 

\subsubsection{Source J012930-733311}
\cite{Murphy2010} identify source J012930-733311 as associated with the HII region IRAS 01283-7349 (NGC 602) in the Small Magellanic Cloud (SMC). A HII region with typical size $\sim$5\,pc can be expected to subtend an angle of $\sim$17 arcsec at the distance of the SMC. The source is compact at 4500\,m baselines of ATCA at 20 GHz (6-km visibility = 0.96 $\pm$0.06), and also shows a visibility profile of an unresolved source at short baselines ($<$ $\sim$ 300m) of ATCA.  It exhibits a flat radio spectrum between 1 and 5 GHz. In the WISE colour-colour plot, the object is found to occupy the central region expected to be occupied by QSOs. The source has WISE extended flag of 0, corresponding to a unresolved source. Thus, based on radio spectra, mid-infrared colours and its unresolved nature in both radio and in WISE catalogue, we conclude that the radio source is a background AGN, with a chance alignment with the SMC region. 

\section{Common names of select sources}
\label{Sec:Appendix-commonNames}
Common names of counterparts for select prominent AT20G sources, whose updated mid-infrared photometry are provided in \autoref{Tab:main_table} are presented in \autoref{Tab:commonNames}. 

\begin{table}
\begin{tabular}{|l|l|l|}
\hline
  \multicolumn{1}{|c|}{AT20G\_name} &
  \multicolumn{1}{c|}{WISEdesig} &
  \multicolumn{1}{c|}{CommonNames} \\
\hline
  J010633-463836 & J010633.01-463836.4 & HIPASS\,J0105-46\\
  J012600-012041 & J012600.61-012042.4 & NGC\,0547\\
  J024104-081520 & J024104.80-081520.7 & NGC\,1052\\
  J070240-284149 & J070240.43-284150.8 & NGC\,2325\\
  J131931-123925 & J131931.67-123924.9 & NGC\,5077\\
  J131949-272437 & J131950.03-272436.8 & NGC\,5078\\
  J132112-434216 & J132112.85-434216.6 & NGC\,5090\\
  J133608-082952 & J133608.26-082951.7 & NGC\,5232\\
  J133639-335756 & J133639.05-335757.0 & IC4296\\
  J145924-164136 & J145924.76-164136.3 & TJ14592472-1641368\\
  J172341-650036 & J172341.04-650036.6 & HIPASS\,J1722-65b\\
  J182507-631453 & J182507.23-631454.0 & NGC\,6614\\
  J194524-552049 & J194524.23-552048.7 & NGC\,6812\\
  J220916-471000 & J220916.21-471000.1 & NGC\,7213\\
  J225710-362744 & J225710.61-362744.0 & HIPASS\,J2257-36\\

\hline

\end{tabular}
\caption{Table presents common names of a small number of AT20G compact sources whose updated WISE counterpart photometry, measured by T.~Jarrett, are presented in \autoref{Tab:main_table}.}
  \label{Tab:commonNames}
  \end{table}

\bsp	
\label{lastpage}
\end{document}